%% file: main.tex
\documentclass[10pt,twocolumn,letterpaper]{article}

\usepackage{iccv}              %
\usepackage[accsupp]{axessibility}  %

\input{packages}
\input{macros}

\definecolor{iccvblue}{rgb}{0.21,0.49,0.74}
\usepackage[pagebackref,breaklinks,colorlinks,allcolors=iccvblue]{hyperref}

\def\paperID{117} %
\def\confName{ICCV}
\def\confYear{2025}

\title{WorldScore: A Unified Evaluation Benchmark for World Generation}

\author{ Haoyi Duan$^{*}$ \quad Hong-Xing Yu$^{*}$ \quad Sirui Chen \quad Li Fei-Fei \quad  Jiajun Wu \vspace{4pt} \\
    Stanford University
}

\begin{document}
\maketitle

\renewcommand{\thefootnote}{\fnsymbol{footnote}}\footnotetext[1]{Equal contribution.}\renewcommand{\thefootnote}{\arabic{footnote}}\setcounter{footnote}{0}

\begin{abstract}

\input{text/0_abstract}
\end{abstract}

\input{text/1_intro}
\input{text/2_rw}
\input{text/3_method}

\input{text/4_exp}
\input{text/5_conclusion}

\myparagraph{Acknowledgments.} This work is in part supported by ONR YIP N00014-24-1-2117, ONR MURI N00014-22-1-2740, NSF RI \#2211258 and \#2338203, and the Okawa Foundation. We thank Mohamed El Banani and Christoph Lassner for their helpful discussion.

{
    \small
    \bibliographystyle{ieeenat_fullname}
    \bibliography{main}
}

\input{supp}

\end{document}

%% file: packages.tex
\usepackage{amsmath,amsfonts,amsthm,amssymb}
\usepackage{mathtools}
\usepackage{bm}
\usepackage{nicefrac}
\usepackage{microtype}
\usepackage{lipsum}
\usepackage{fontawesome}
\usepackage{tcolorbox}
\usepackage{xcolor}
\usepackage{tabularx}

\usepackage{color,xcolor}
\usepackage{epsfig}
\usepackage{graphicx}

\usepackage{adjustbox}
\usepackage{array}
\usepackage{booktabs}
\usepackage{colortbl}
\usepackage{wrapfig}
\usepackage{hhline}
\usepackage{multirow}
\usepackage{subcaption}
\usepackage[size=small]{caption}
\usepackage{float}
\usepackage{pgfmath}

\usepackage{changepage}
\usepackage{extramarks}
\usepackage{fancyhdr}
\usepackage{lastpage}
\usepackage{setspace}
\usepackage{soul}
\usepackage{xspace}

\usepackage{url}

\usepackage{algorithm}
\usepackage{algorithmicx}
\usepackage{algpseudocode}
\usepackage{enumerate}
\usepackage{enumitem}  %
\usepackage{makecell}
\usepackage{pifont}
\usepackage{titlecaps}
\usepackage[accsupp]{axessibility}
\usepackage{framed}

%% file: macros.tex
\newcommand{\modelfull}{WorldScore\xspace}
\newcommand{\benchmark}{WorldScore\xspace}
\newcommand{\metric}{WorldScore\xspace}
\newcommand{\wss}{WorldScore-Static\xspace}
\newcommand{\wsd}{WorldScore-Dynamic\xspace}

\newcommand{\website}{\url{https://haoyi-duan.github.io/WorldScore/}}

\definecolor{darkgreen}{rgb}{0.29, 0.64, 0.35} 
\newcommand{\cmark}{\textcolor{green}{\ding{51}}} %
\newcommand{\xmark}{\textcolor{red}{\ding{55}}} %

\newcommand{\myparagraph}[1]{\vspace{0.1cm}\noindent\textbf{#1}}
\renewcommand{\paragraph}[1]{\vspace{0.1cm}\noindent\textbf{#1}}

\definecolor{MyDarkBlue}{rgb}{0,0.08,1}
\definecolor{MyAqua}{rgb}{0,0.7,0.7}
\definecolor{MyDarkGreen}{rgb}{0.02,0.6,0.02}
\definecolor{MyDarkRed}{rgb}{0.8,0.02,0.02}
\definecolor{MyDarkOrange}{rgb}{0.40,0.2,0.02}
\definecolor{MyPurple}{RGB}{111,0,255}
\definecolor{MyRed}{rgb}{1.0,0.0,0.0}
\definecolor{MyGold}{rgb}{0.75,0.6,0.12}
\definecolor{MyDarkgray}{rgb}{0.66, 0.66, 0.66}

\definecolor{Cardinal}{rgb}{0.549,0.082,0.082}

%% file: text/0_abstract.tex
We introduce the \benchmark benchmark, the first unified benchmark for world generation. We decompose world generation into a sequence of next-scene generation tasks with explicit camera trajectory-based layout specifications, enabling unified evaluation of diverse approaches from 3D and 4D scene generation to video generation models. The \benchmark benchmark encompasses a curated dataset of 3,000 test examples that span diverse worlds: static and dynamic, indoor and outdoor, photorealistic and stylized. The \metric metric evaluates generated worlds through three key aspects: controllability, quality, and dynamics. Through extensive evaluation of 20 representative models, including both open-source and closed-source ones, we reveal key insights and challenges for each category of models.
Our dataset, evaluation code, and leaderboard can be found at \website.

%% file: text/1_intro.tex
\input{fig_text/comparison_vbench}

\input{fig_text/benchmark_comparison}

\section{Introduction}

Recent advances in visual generation have sparked growing interest in world generation---the creation of large-scale, diverse worlds with various scenes, which finds wide applications in entertainment, education, simulation, and embodied AI. The rapid progress in video generation~\cite{agarwal2025cosmos, videoworldsimulators2024sora, yang2024cogvideox, chen2024videocrafter2}, 3D scene generation~\cite{fridman2024scenescape, yu2024wonderjourney, chung2023luciddreamer, yu2024wonderworld}, and 4D scene generation~\cite{bahmani20244dfy, yu20244real, xu2024comp4d} has shown generating high-quality individual scenes, demonstrating the potential of these models as world generation systems. However, as the concept of world generation expands, users demand to generate more comprehensive worlds that seamlessly integrate multiple varied scenes with detailed spatial layout controls rather than disconnected individual environments.

Achieving this vision requires a unified evaluation benchmark that systematically assesses different types of world generation models across large-scale, diverse worlds, which is currently absent. Existing benchmarks mainly focus on video generation~\cite{feng2024tcbench, liu2024evalcrafter, liu2024fetv, meng2024towardsworldsimulator, yuan2024chronomagic} and evaluate only individual scene generation. For example, VBench~\citep{huang2024vbench} primarily evaluates text-to-video (T2V) tasks using curated prompts without explicit spatial layout control, restricting their evaluations to single scenes (Figure~\ref{fig:comparison_vbench}). Moreover, despite the promising potential of 3D and 4D scene generation methods for world generation, current benchmarks lack essential components such as camera specifications and reference images, making them incompatible with many state-of-the-art 3D/4D scene generation methods that require an image or a camera trajectory as inputs~\citep{fridman2024scenescape, yu2024wonderjourney, chung2023luciddreamer, yu2024wonderworld,lee2024vividdream}.

We introduce \benchmark, a unified benchmark for world generation. Our key design is to decompose world generation into a sequence of next-scene generation tasks, where each step is characterized by a triplet of \texttt{(current scene, next scene, layout)}. For unified evaluation across different methods, we provide both an image and a text prompt for a \texttt{current scene}, as well as both camera matrices and a textual description for a \texttt{layout} specification. This design allows our \benchmark benchmark to evaluate various approaches including 3D, 4D, text-to-video, and image-to-video models on large-scale world generation. All methods are evaluated on a common output format, i.e., rendered or generated videos, to enable direct comparison of generation across different types of approaches.

Our evaluation metric, \metric, is computed by aggregating three key aspects: \emph{controllability}, which measures the adherence of the generated worlds w.r.t. control inputs; \emph{quality}, which measures the fidelity and consistency; \emph{dynamics}, which measures how much the generated worlds exhibit accurate and stable motions. Each of these aspects comprises a few distinct metrics, leading to a total of 10 metrics that contribute to computing the \metric. 

To enable a comprehensive assessment, we curate a diverse dataset covering both static and dynamic world generation scenarios across different visual domains. For static worlds, we include 5 categories of indoor scenes and 5 categories of outdoor scenes with varying sequence lengths. For dynamic worlds, we include 5 distinct types of dynamics such as rigid motion and fluid motion. Additionally, each example in our dataset has a corresponding stylized counterpart sampled from a rich set of candidate styles, allowing the evaluation of various visual domains. In total, our dataset comprises 3000 high-quality test examples that span indoor/outdoor environments and photorealistic/stylized visual domains.

We conduct extensive experiments by evaluating 20 diverse models, including 6 image-to-video models (with 2 leading closed-source models), 7 text-to-video models, 6 3D scene generation models, and a 4D generation model. 
In summary, our contributions are fourfold:
\begin{itemize}
\item We propose the first world generation benchmark, \benchmark, which allows unified evaluation across various approaches including 3D, 4D, I2V, and T2V models. 
\item We curate a high-quality, diverse dataset for our benchmark evaluation. Our dataset covers diverse static and dynamic scenes across various categories with multiple visual styles.
\item We introduce the \metric metrics, which aggregate critical aspects in world generation model performances, including controllability, quality, and dynamics.
\item Through the comprehensive evaluation of 18 open-source and 2 closed-source models, we reveal key insights and challenges in current world generation approaches, providing valuable guidance for future research.
\end{itemize}

%% file: fig_text/comparison_vbench.tex
\begin{figure}
    \centering
    \includegraphics[width=0.45\textwidth]{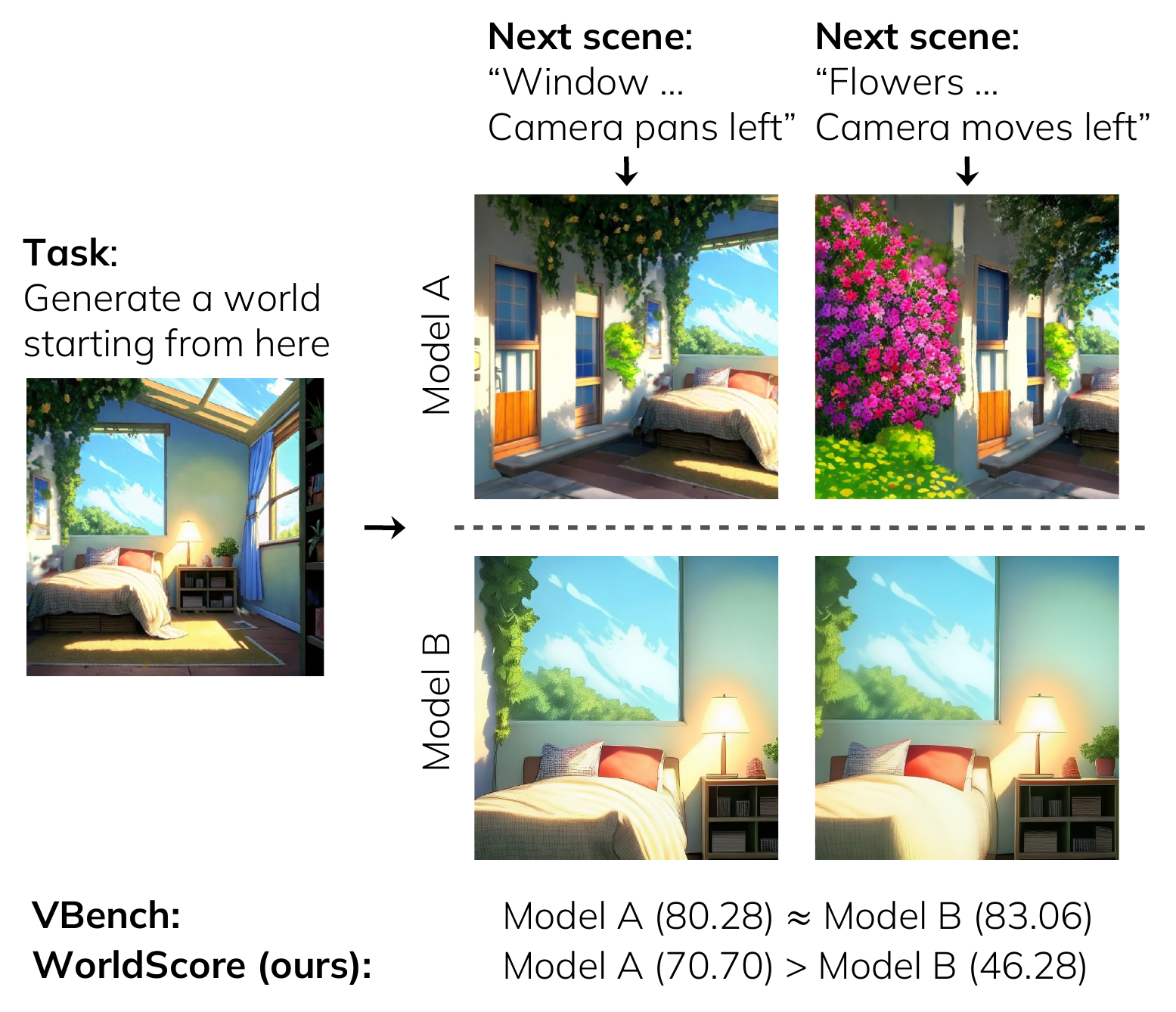}
    \vspace{-0.1cm}
    \caption{While existing video benchmarks like VBench~\citep{huang2024vbench} rate Models A and B similarly based on single-scene video quality, our \benchmark benchmark differentiates their world generation capabilities by identifying that Model B fails to generate a new scene or follow the instructed camera movement. In \website, we show the videos to explain our \metric metrics.
    } 
    \label{fig:comparison_vbench}
    \vspace{-0.2cm}
\end{figure}

%% file: fig_text/benchmark_comparison.tex
\begin{table*}[t]
\resizebox{\textwidth}{!}{%
\centering
\small
\setlength{\tabcolsep}{4.5pt} 
\begin{tabular}{lccccccccc}
\toprule
Benchmark & \# Examples & Multi-Scene & Unified & Long Seq. & Image Cond. & Multi-Style & Camera Ctrl. & 3D Consist. \\
\midrule
TC-Bench~\cite{feng2024tcbench}       & 150  & \xmark & \xmark & \xmark & \cmark & \xmark & \xmark & \xmark \\
EvalCrafter~\cite{liu2024evalcrafter}  & 700  & \xmark & \xmark & \xmark & \xmark & \xmark & \xmark & \xmark \\
FETV~\cite{liu2024fetv}                & 619  & \xmark & \xmark & \xmark & \xmark & \xmark & \xmark & \xmark \\
VBench~\cite{huang2024vbench}           & 800  & \xmark & \xmark & \xmark & \xmark & \xmark & \xmark & \xmark \\
T2V-CompBench~\cite{sun2024t2vcompbench} & 700 & \xmark & \xmark & \xmark & \xmark & \xmark & \xmark & \xmark \\
Meng et al.~\cite{meng2024towardsworldsimulator} & 160  & \xmark & \xmark & \xmark & \xmark & \xmark & \xmark & \xmark \\
Wang et al.~\cite{wang2024yoursimulatorgoodstorypresenter} & 423  & \xmark & \xmark & \cmark & \xmark & \xmark & \xmark & \xmark \\
ChronoMagic-Bench~\cite{yuan2024chronomagic}  & 1649 & \xmark & \xmark & \xmark & \xmark & \xmark & \xmark & \xmark \\
WorldModelBench~\cite{li2025worldmodelbench}  & 350 & \xmark & \xmark & \xmark & \cmark & \xmark & \xmark & \xmark \\
\textbf{\modelfull (Ours)} & \textbf{3000} & \cmark & \cmark & \cmark & \cmark & \cmark & \cmark & \cmark \\
\bottomrule
\end{tabular}
}
\vspace{-0.1cm}
\caption{\textbf{Comparison of Benchmarks.} Our \benchmark benchmark is designed to evaluate various world generation approaches including 3D, 4D, I2V and T2V models. It is designed to generate multiple scenes with varying sequence lengths. Our benchmark also features multiple visual styles, accurate camera control evaluation, and 3D consistency evaluation, all of which are important factors in world generation yet currently missing in existing benchmarks.}
\vspace{-0.3cm}
\label{tab:benchmark_comparison}
\end{table*}

%% file: text/2_rw.tex
\section{Related Work}

\input{fig_text/framework}

\myparagraph{Video generation benchmarks.}
The progress of both open-source~\cite{yang2024cogvideox, xing2023dynamicrafter, chen2024videocrafter2, agarwal2025cosmos} and closed-source~\cite{videoworldsimulators2024sora, gen3, minimax,luma} video generation models has stimulated the proposal of numerous benchmarks~\cite{feng2024tcbench, huang2024vbench, liu2024evalcrafter, liu2024fetv, meng2024towardsworldsimulator, yuan2024chronomagic}. However, most existing benchmarks, such as VBench~\cite{huang2024vbench} and WorldModelBench~\citep{li2025worldmodelbench}, focus on evaluating video generation models based on single-scene video quality without layout control and multi-scene generation. Furthermore, their designs are incompatible with 3D/4D scene generation methods that require camera specification. In contrast, our \benchmark benchmark is designed to focus on evaluating world generation approaches with multi-scene generation tasks, and it is designed to accommodate 3D, 4D, I2V and T2V models. We show a detailed comparison in Table~\ref{tab:benchmark_comparison}.

\myparagraph{Video generation models.} Recent advances in image generation, including VAEs~\cite{kingma2013autoencoder}, GANs~\cite{goodfellow2014generativegan, mirza2014conditionalgan, brock2018largescalegan, karras2017progressivegan, karras2019stylegan, karras2020analyzingstylegan, karras2021aliasgan}, VQ approaches~\cite{van2017neuralvq, esser2021taming}, and Diffusion models~\cite{ho2020denoisingdiffusion, sohl2015deepnonequilibrium, song2020scorebased, peebles2023scalablediffusionmodels}, have fueled explorations in video generation~\cite{hong2022cogvideo, luo2023videofusion, wang2024lavie, wang2023modelscope, singer2022make}. The advent of Sora~\cite{videoworldsimulators2024sora} has further demonstrated the potential of video models as world generation models~\cite{kang2024farhowfar, meng2024towardsworldsimulator, xiang2024pandora}. While most recent models focus on text-to-video (T2V) generation~\cite{chen2023videocrafter1, chen2024videocrafter2, fan2025vchitect, li2024t2v}, developments in image-to-video (I2V)~\cite{zhou2024allegro, xu2024easyanimate, yang2024cogvideox, xing2023dynamicrafter, agarwal2025cosmos} have also been significant. In our \benchmark benchmark, we evaluate both T2V and I2V models as world generation approaches, thanks to our unified design that accommodates both image and text conditioning strategies.

\myparagraph{3D scene generation.} Besides video models, our \benchmark benchmark also includes 3D and 4D generation methods. Recent 3D scene generation models rely mainly on generative diffusion models~\cite{fridman2024scenescape, yu2024wonderjourney}, which formulate generating scenes in a sequential manner using supervision from 2D image outpainting models. These methods~\cite{hollein2023text2room, chung2023luciddreamer, engstler2024invisible, yu2024wonderworld} project the synthesized 2D scene extensions into a 3D representation by leveraging depth estimation models~\cite{lasinger2019towardsmidas, bhat2023zoedepth, yang2024depthanything, ke2024repurposingmarigold}.

To incorporate dynamics, 4D generation methods~\cite{ling2024align4d, zheng2024unified, ren2023dreamgaussian4d, zhao2023animate124, zheng2024unified, singer2023textto4d,lee2024vividdream} further integrate multi-view and video diffusion priors. Due to the difficulty of scene-level generation, most of existing methods focus on object-level generation. Nevertheless, we include 4D-fy~\cite{bahmani20244dfy} in our benchmark due to its open-source accessibility.

%% file: fig_text/framework.tex
\begin{figure*}[ht]
    \centering
    \includegraphics[width=0.9\textwidth]{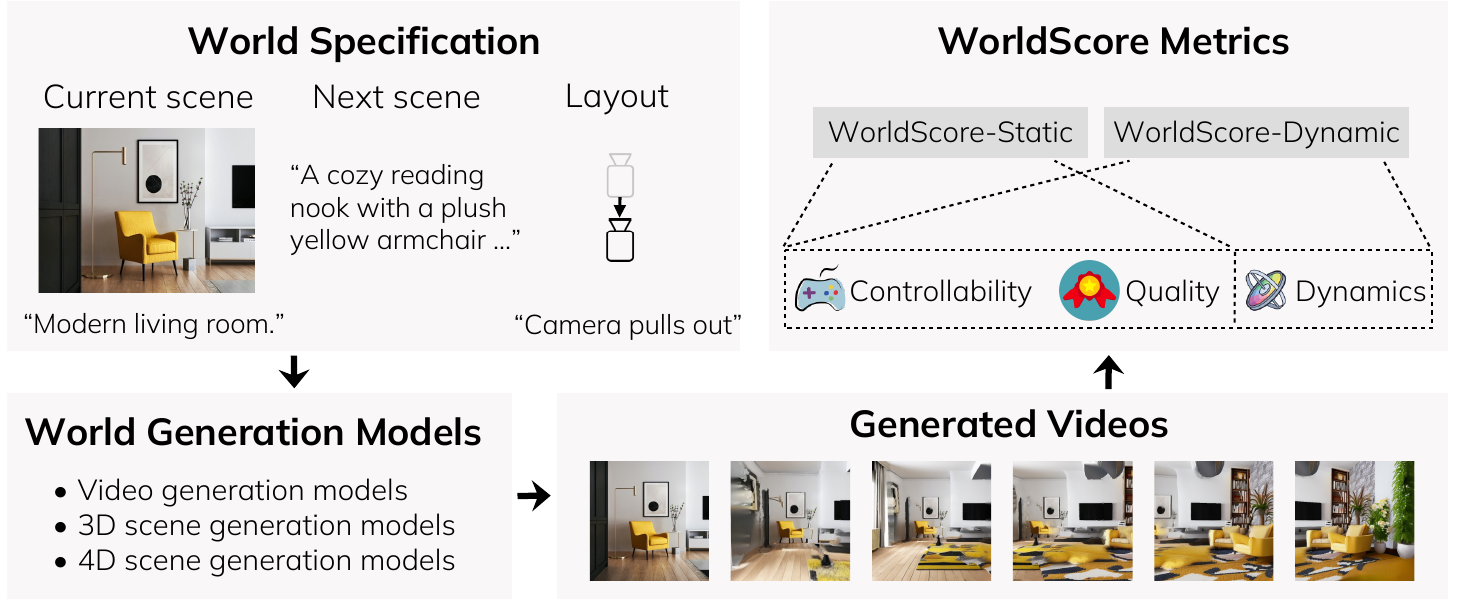}
    \vspace{-0.1cm}
    \caption{\textbf{Overview of the \benchmark benchmark design.} \emph{Top left:} World generation is decomposed into a sequence of next-scene generation tasks, where each step follows a structured world specification defining both spatial layout and semantic content. \emph{Bottom left:} The unified world specification is used to instruct different types of models, including video generation and 3D/4D generation models. \emph{Bottom right:} All models output videos for evaluation. \emph{Top right:} Output videos are evaluated using the \metric metrics, which assess three fundamental aspects including controllability, quality, and dynamics.}
    \vspace{-0.2cm}
    \label{fig:framework}
\end{figure*}

%% file: text/3_method.tex
\input{fig_text/dataset_overview}

\section{The \benchmark Benchmark}

\myparagraph{Design overview.}
Our goal is to establish an evaluation benchmark for world generation that unifies different methodological approaches. Our \benchmark benchmark introduces three key components: (1) a standardized world specification, (2) a carefully curated dataset, and (3) multi-faceted metrics. We show an overview in Figure~\ref{fig:framework}. We decompose world generation into a sequence of next-scene generation tasks, where each step is defined by a world specification encompassing both spatial layout and semantic content (top left of Figure~\ref{fig:framework}). This world specification enables us to instruct different types of models ranging from 3D/4D scene generation to video generation approaches. The generated outputs, standardized as videos (bottom right of Figure~\ref{fig:framework}), are then evaluated using the \metric metrics (top right of Figure~\ref{fig:framework}) that assess three critical aspects: controllability, quality, and dynamics. This unified evaluation approach ensures fair comparison across different methodological paradigms.

\subsection{World Specification}
\myparagraph{Formulation.} We decompose the world generation task into a sequence of next-scene generation tasks, where each step is specified by a triplet of $(\mathcal{C}, \mathcal{N}, \mathcal{L})$, where $\mathcal{C}=\{\mathbf{I},\mathcal{P}\}$ denotes the current scene given by a scene image $\mathbf{I}$ and a text prompt $\mathcal{P}$, $\mathcal{N}$ denotes the next-scene text prompt, and $\mathcal{L}=\{\mathcal{T},\mathcal{Y}\}$ denotes the layout given by a camera trajectory  $\mathcal{T}=(\mathbf{C}_1,\mathbf{C}_2,\cdots,\mathbf{C}_N)$ where $\mathbf{C}_i$ denotes a camera matrix and a text prompt of camera movement $\mathcal{Y}$. Then, a world generation model is instructed to generate a video:
\begin{equation}
    \mathbf{V} = g_\text{world}(w_\text{proc}(\mathcal{C}, \mathcal{N}, \mathcal{L})),
\label{eq:world_generation}    
\end{equation}
where $\mathbf{V}$ denotes a video, $g_\text{world}$ denotes the world generation model, and $w_\text{proc}$ denotes a model-specific pre-processing which we detail in Supp.~\ref{supp_sec:supplementary_world_specification}.

\myparagraph{Static and dynamic worlds.} We explicitly disentangle the evaluation of dynamics aspect from the controllability and quality aspects due to their distinct natures. To this end, we have two types of tasks: 

\vspace{0.1cm}
\noindent
\underline{Static world generation}: We instruct a model to generate varying-length scene sequences for controllability and quality assessment. Here, the next-scene text prompt $\mathcal{N}$ describes the new scene contents, and the layout $\mathcal{L}$ describes large camera movements.

\vspace{0.1cm}
\noindent
\underline{Dynamic world generation}: We instruct a model to generate in-scene motion for dynamics assessment. Here, the next-scene text prompt $\mathcal{N}$ describes the same scene content as $\mathcal{C}$ but with dynamics changes, e.g., an animal moving. The layout $\mathcal{L}$ explicitly specifies a fixed camera position without any camera motion.

\input{fig_text/dataset_curation}

\subsection{Dataset Curation}
Our dataset consists of 3000 examples (world specifications), including 2000 for static world generation and 1000 for dynamic world generation. We show a detailed statistics in Table~\ref{tab:dataset_stats} in the supplementary material.

\myparagraph{Curation on current scene $\mathcal{C}$.} The current scene $\mathcal{C}=\{\mathbf{I},\mathcal{P}\}$ is given by an image $\mathbf{I}$ and its text prompt $\mathcal{P}$. We show an illustration of our curation process in Figure~\ref{fig:image_suite}.

For static world generation, we define 10 categories of scenes including 5 indoor and 5 outdoor scene types. Then, we source images from open-source scene datasets~\cite{zhu2022learninginterviorverse, roberts2021hypersim, song2015sun, Matterport3D, vasiljevic2019diode, eth3d, lhq, le2021eden, argoverse} and supplement with an online source, Unsplash~\cite{unsplash}. We apply a very rigorous filtering strategy to ensure high quality and high diversity (Supp.~\ref{sec:filtering}), leading to approximately 5000 images $\mathbf{I}$ in photorealistic style (they are either real photos or physically-based rendered images). Then, we query a Vision-Language Model (VLM), GPT-4o~\cite{gpt4o}, to generate captions $\mathcal{P}$ for these images and do a 10-way classification to put each of them into a category. Finally, we further filter each category by keeping the first 100 highest-quality images, leading to 1000 images $\mathbf{I}$ and their corresponding prompts $\mathcal{P}$.

Then, we create a stylized counterpart for each example in the photorealistic domain. For each example, we randomly pick a style from a set of 7 style candidates, and create a new text prompt $\mathcal{P}$ by adding the style text to the prompt of the photorealistic example (Supp.~\ref{sec:stylized}). Then, we leverage a commercial style-controlled text-to-image generation model~\citep{recraft} to generate the stylized counterpart image $\mathbf{I}$. We show examples in the top two rows in Figure~\ref{fig:dataset_overview}.

For dynamic world generation, we define 5 categories of motion types and source Unsplash to manually curate 100 images for each of the category. We follow a similar process as in the static world generation examples to create text prompts and stylized counterpart, eventually leading to a total of 1000 examples. We show examples in the bottom row in Figure~\ref{fig:dataset_overview}.

\input{fig_text/layout}

\myparagraph{Curation on next-scene text prompts $\mathcal{N}$.}
Each world generation consists of a sequence of next-scene generation tasks. The next-scene text prompt $\mathcal{N}$ can have varying lengths. In particular, we consider two cases: (1) a small world where $\mathcal{N}$ consists of only one new scene, and (2) a large world where $\mathcal{N}$ consists of three new scenes.

To generate coherent and contextually relevant scene sequences, we adopt an auto-regressive scene description generation process~\cite{yu2024wonderjourney}, that is, we instruct an LLM to generate the next-scene text prompt that should be different from all current scene text prompts. For example, for a small world,
\begin{equation}
    \mathcal{N} = \text{LLM}(\mathcal{J}, \mathcal{P}),
\end{equation}
where the LLM takes two inputs: (1) the task specification $\mathcal{J}$ = \textit{``Generate a scene description different from the past scenes.''}
\footnote{This is a brief summary of the actual prompt provided in Supp.~\ref{supp_sec:prompt_suit_details}.}, and (2) a collection of past and current scene descriptions. For a large world which consists of 4 scenes, we repeat this process for 3 times, so that $\mathcal{N}=\mathcal{N}_1+\mathcal{N}_2+\mathcal{N}_3$ consists of three individual next-scene prompts. In our generation, 20\% of our static world generation examples are large worlds, and the others are small worlds.

\myparagraph{Curation on layouts $\mathcal{L}$.} 
A layout $\mathcal{L}=\{\mathcal{T},\mathcal{Y}\}$ is given by a camera trajectory $\mathcal{T}=(\mathbf{C}_1,\mathbf{C}_2,\cdots,\mathbf{C}_N)$ and a text prompt of camera movement $\mathcal{Y}$. 
We curate a set of 8 camera movements (left of Figure~\ref{fig:layout}) which are widely used in movie industry. This design achieves two objectives: Firstly, it covers all spatial directions; secondly, it facilitates text-to-video models to take the instruction $\mathcal{Y}$ as most of them are trained on movie clips that often contain these camera movement descriptions. These movements include both intra-scene movements, such as moving into a scene, as well as inter-scene transitions, such as pulling out the camera. For each static scene generation example, we randomly assign a layout $\mathcal{L}$ to a next-scene generation task. We show an example in the right of Figure~\ref{fig:layout}. When the assigned layout is intra-scene, we perform a replacement of $\mathcal{N}$ with $\mathcal{P}$.

We leave details of our dataset curation in Supp.~\ref{supp_dataset}.

\input{fig_text/object_control}

\subsection{The \metric Metrics}

Our \metric metrics include two overall scores: \wss which measures only the static world generation capability, and \wsd which measures dynamic world generation capability in addition to static worlds. They are defined as the aggregation of several individual metrics in the three key aspects: controllability, quality, and dynamics. We briefly introduce each individual metric in the following, and we leave details in Supp.~\ref{supp_sec:supplementary_of_worldscore_eval}.

\myparagraph{Controllability.} We have three metrics.

\vspace{0.1cm}
\noindent
\underline{Camera controllability}: To evaluate how the models adhere to the instructed layout $\mathcal{L}=\{\mathcal{T},\mathcal{Y}\}$, we compute camera errors as follows:
\begin{equation}
    e_{\text{camera}} = \sqrt{e_{\theta} \cdot e_t},
    \label{algorithm:camera_error_geometric_mean}
\end{equation}
where $e_{\theta}$ and $e_t$ are scale-invariant rotation and translation errors with respect to the ground truth trajectory $\mathcal{T}$, respectively. We compute camera errors across all the frames of the generated video $\mathbf{V}$. We leave more details in Supp.~\ref{supp_sec:camera_control}.

\vspace{0.1cm}
\noindent
\underline{Object controllability}: We evaluate whether the objects specified in the next-scene prompt $\mathcal{N}$ appear in the generated next scene. To this end, we measure the success rate of object detection. Specifically, we leverage a state-of-the-art open-set object detection model~\citep{liu2025grounding}. We extract one or two individual object descriptions from the text prompt $\mathcal{N}$. We compute the success rate by matching the detected objects with the object descriptions. This provides a quantitative measure of how well the generated foreground objects adheres to the world specification.

\vspace{0.1cm}
\noindent
\underline{Content alignment}: Besides the objects (which typically occupies approximately only $\frac{1}{4}$ of the text prompt length), we also assess whether the generated scenes are aligned with the entire text $\mathcal{N}$ using CLIPScore \cite{hessel2021clipscore}.

\myparagraph{Quality.} We have four metrics.

\vspace{0.1cm}
\noindent
\underline{3D consistency}: We evaluate the 3D consistency in the static world videos. This metric focuses on how the geometry of a scene remains stable across frames, regardless of slight changes in visual textures. To this end, we use DROID-SLAM~\citep{teed2021droid}, a standard SLAM method, to estimate dense pixel-wise depth for each frame, and then we compute the reprojection error between a pair of co-visible pixels in consecutive frames. Since DROID-SLAM is designed to be robust against appearance changes, this metric measures geometric inconsistency. We show an example in Figure~\ref{fig:object_control}, and we leave more details in Supp.~\ref{supp_sec:3d_consistency}.

\vspace{0.1cm}
\noindent
\underline{Photometric consistency}: While 3D consistency exclusively focuses on geometry, photometric consistency focuses on appearance (e.g., textures). Many video generation models struggle with maintaining consistent object textures, leading to appearance inconsistency issues such as texture flickering. Existing consistency metrics, such as those with CLIP or DINO features \cite{huang2024vbench, huang2024vbench++}, focus on categorical identity but fail to capture fine-grained texture changes. For example, the mountain in the middle row of Figure~\ref{fig:object_control} remains a mountain (i.e., the same geometry and semantic class) across frames, but the texture (grass) has been shifted and distorted over time. This cannot be captured by CLIP/DINO features. 

To detect photometric artifacts, our photometric consistency metric estimates the optical flow between consecutive frames and computes the Average End-Point Error (AEPE). This metric effectively identifies unstable visual appearance, as shown in Figure~\ref{fig:object_control}. We leave more details in Supp.~\ref{supp_sec:photometric_consistency}.

\vspace{0.1cm}
\noindent
\underline{Style consistency}: We evaluate the style consistency by computing the differences (F-norm) between the Gram matrices~\citep{gatys2015neuralgrammatrix} of the first frame and the last frame of a single next-scene generation task.

\vspace{0.1cm}
\noindent
\underline{Subjective quality}: We use automatic metrics to evaluate the human perceptual quality of the generated scenes. There exists some automatic image assessment metrics~\citep{qalign} and aesthetic metrics~\citep{wang2023exploringclipiqa}, and thus we consider ensemble them. To find a combination that best fits human perception, we perform a human study of 400 participants, enumerate different metric combinations, and we pick the combination (CLIP-IQA+ \cite{wang2023exploringclipiqa} with CLIP Aesthetic \cite{schuhmann2022clip+aesthetic}) that best matches human preference. We leave more details in Supp.~\ref{supp_sec:subjective_quality}.

\input{fig_text/evaluation_results_all}

\myparagraph{Dynamics.} We have three metrics.

\vspace{0.1cm}
\noindent
\underline{Motion accuracy}: Accurate motion placement is essential in dynamics generation. For example, if a prompt specifies that a car should move while nearby pedestrians remain still, the model should animate the car, not the pedestrians. To quantify this, we introduce motion accuracy, which measures whether the motion specified in the next-scene prompt $\mathcal{N}$ occurs in the designated regions. As shown in the bottom row of Figure~\ref{fig:object_control}, the score is calculated by comparing optical flow within the intended region with the flow outside the region. We need to consider the outside flow as it cancels out the global motion caused by unintended camera movements.

\vspace{0.1cm}
\noindent
\underline{Motion magnitude}: We measure a world generation model's ability to create large motions by estimating the optical flow between the consecutive frames of the generated video.

\vspace{0.1cm}
\noindent
\underline{Motion smoothness}: Temporal jittering is a common failure mode in dynamic world generation. We utilize a standard video frame interpolation model~\citep{zhang2024vfimamba} to generate smooth interpolation as ground truth to evaluate the temporal smoothness of generated videos $\mathbf{V}$. We leave details in Supp.~\ref{supp_sec:motion_smoothness}.

\myparagraph{Score normalization and aggregation.} After computing individual evaluation metrics, we apply a linear normalization and mapping process based on empirical bounds (Supp.~\ref{supp_sec:empirical_bounds}) to ensure that the final scores fall within the range between zero to one, and then we scale it by 100. Then, we compute the arithmetic mean of the dimension scores within control and quality aspects to obtain our \textbf{\wss}. Additionally, we further incorporate three dynamics dimension scores into the aggregation, resulting in \textbf{\wsd}. For 3D scene generation models that do not support dynamic tasks, we assign 0 to each dynamics metric.

%% file: fig_text/dataset_overview.tex
\begin{figure*}[ht]
    \centering
    \includegraphics[width=0.97\textwidth]{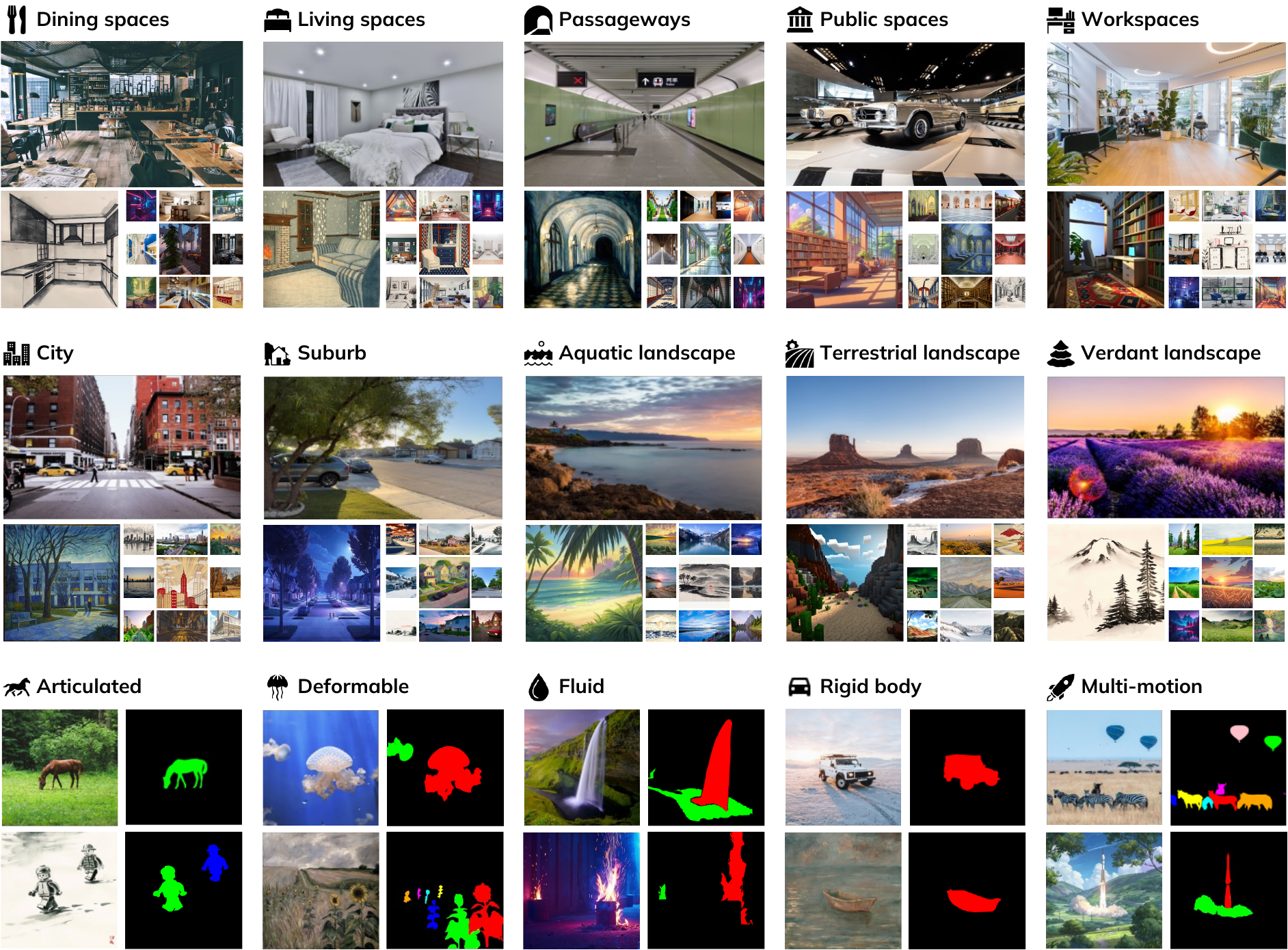}
    \vspace{-0.1cm}
    \caption{\textbf{Showcasing of the current scene images.} \textit{Top two rows:} Static world generation examples are categorized into indoor (first row) and outdoor (second row) scenes, each containing 5 categories. \textit{Bottom row:} Dynamic world generation examples are divided into 5 motion types. Each dynamic example comes with an annotation of motion mask that indicates where the motion should happen.}
    \vspace{-0.3cm}
    \label{fig:dataset_overview}
\end{figure*}

%% file: fig_text/dataset_curation.tex
\begin{figure}[t]
    \centering
    \includegraphics[width=0.48\textwidth]{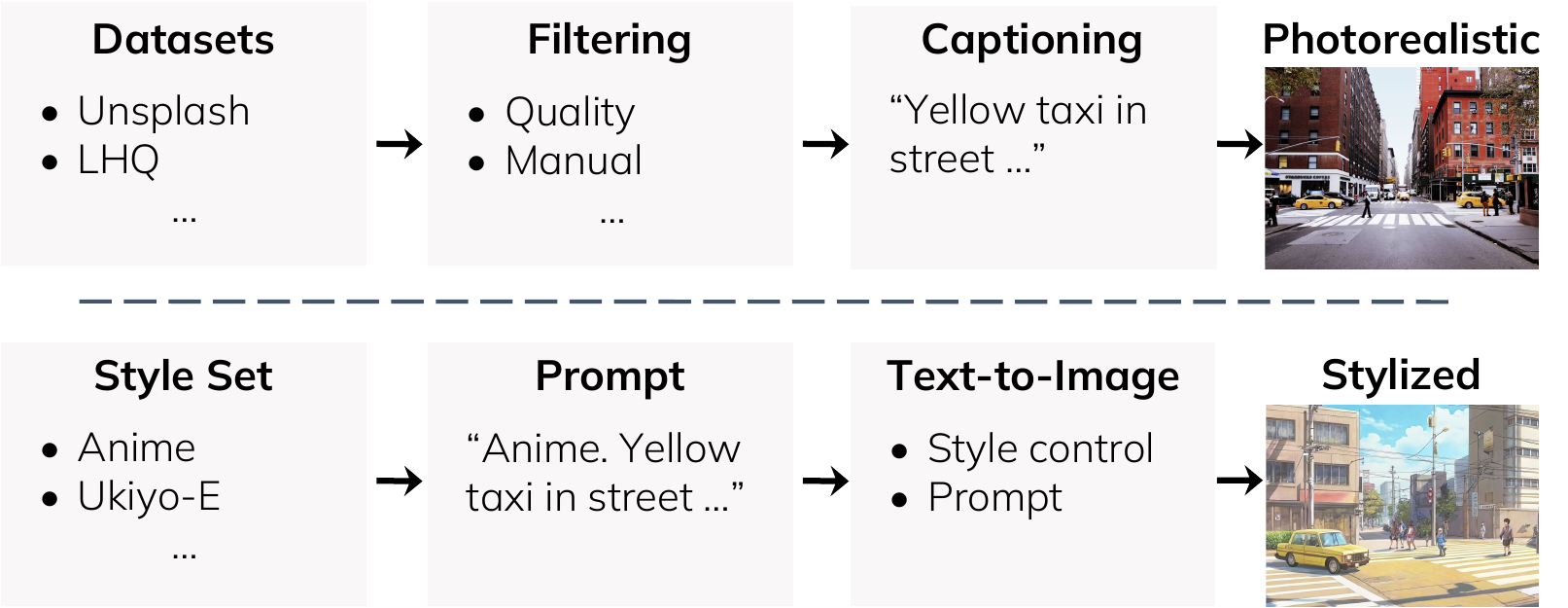}
    \vspace{-0.3cm}
    \caption{\textbf{Curation on the current scene $\mathcal{C}$.} \textit{Top:} Photorealistic worlds. \textit{Bottom:} Stylized counterparts.}
    \vspace{-0.3cm}
    \label{fig:image_suite}
\end{figure}

%% file: fig_text/layout.tex
\begin{figure}
    \centering
    \includegraphics[width=\linewidth]{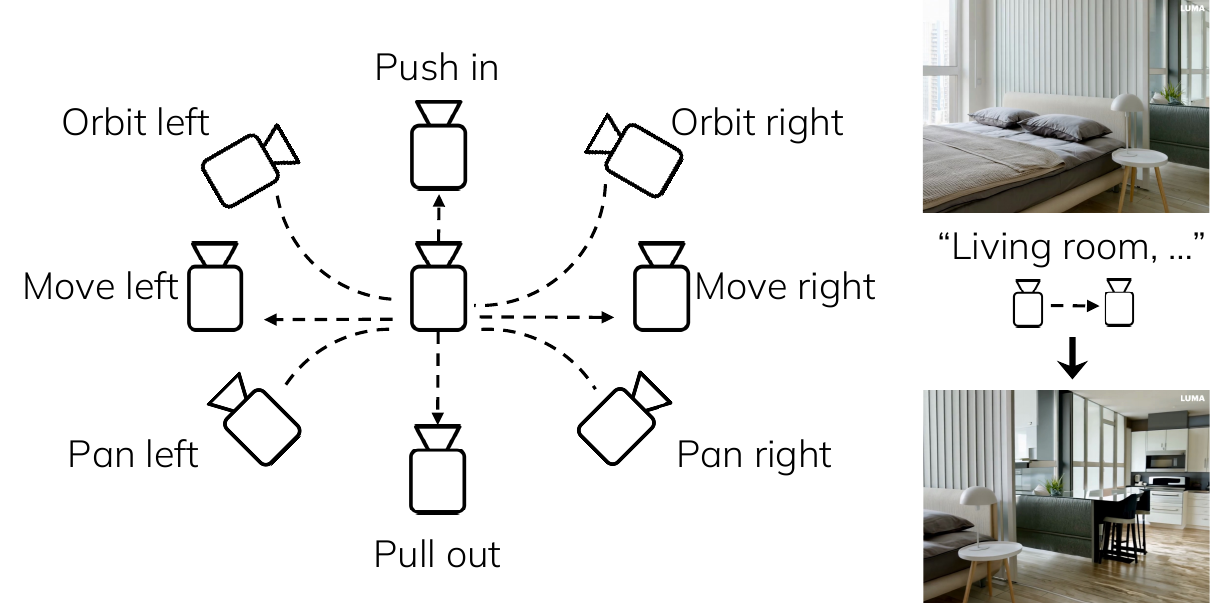}
    \vspace{-0.3cm}
    \caption{\textbf{Curation on layouts $\mathcal{L}$.} \textit{Left:} Camera paths $\mathcal{T}$ and text $\mathcal{Y}$. \textit{Right:} A move-right example.} 
    \vspace{-0.3cm}
    \label{fig:layout}
\end{figure}

%% file: fig_text/object_control.tex
\begin{figure*}
    \centering
    \includegraphics[width=0.9\textwidth]{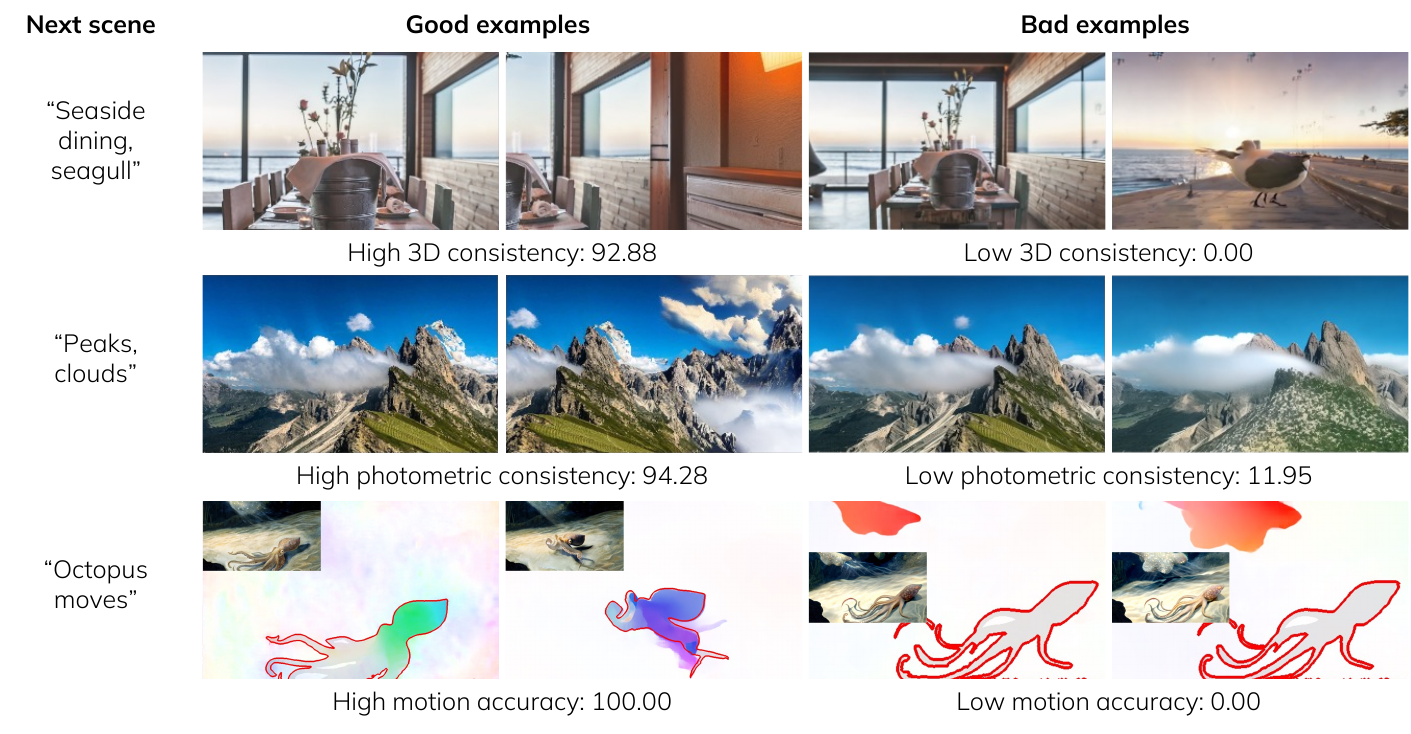}
    \vspace{-0.3cm}
    \caption{\textbf{Typical examples.} \emph{Top: 3D consistency.}  The bad example on the right-hand-side has a sudden change in geometry rather than smooth transition. \emph{Middle:  Photometric consistency.} The bad example exhibits severe texture shift in the mountain grassland. \emph{Bottom: Motion accuracy.}  In the good example, the octopus moves while the jellyfish remains static. For bad example on the right, the jellyfish moves while the octopus remains static. A full version of all metrics is in Figure~\ref{supp_fig:metrics_bigfigure} and Figure~\ref{supp_fig:metrics_bigfigure_dynamics} in supplementary material. In \website, we show videos to explain our \metric metrics.}
    \vspace{-0.3cm}
    \label{fig:object_control}
\end{figure*}

%% file: fig_text/evaluation_results_all.tex
\begin{table*}[t]
\resizebox{\textwidth}{!}{%
\centering
\small 
\setlength{\tabcolsep}{3pt} 

\begin{tabular}{@{}l*{12}{c}@{}} 
\toprule
\multirow{3}{*}{Models} & \multicolumn{2}{c}{WorldScore} & 
\multicolumn{3}{c}{Controllability} & 
\multicolumn{4}{c}{Quality} &
\multicolumn{3}{c}{Dynamics} \\ \cmidrule(lr){2-3}
\cmidrule(lr){4-6} \cmidrule(lr){7-10} \cmidrule(lr){11-13} 
& \rotcell{-Static} & \rotcell{-Dynamic} &
\rotcell{Camera\\Ctrl} & 
\rotcell{Object\\Ctrl} & 
\rotcell{Content\\Align} & 
\rotcell{3D\\Consist} & 
\rotcell{Photo\\Consist} & 
\rotcell{Style\\Consist} &
\rotcell{Subjective\\Qual} &
\rotcell{Motion\\Acc} & 
\rotcell{Motion\\Mag} & 
\rotcell{Motion\\Smooth} \\ 
\midrule
Gen-3~\cite{gen3}          &    {60.71}     & \ul{57.58} & {29.47}  & \ul{62.92} & {50.49} & {68.31} & {87.09} & {62.82} & {63.85} & {54.53}  & {27.48} & {68.87} \\
Hailuo~\cite{minimax}      &    {57.55}        & {56.36} & {22.39}  & \textbf{69.56} & \ul{73.53} & {67.18} & {62.82} & {54.91} & {52.44} & {63.46}  & {27.20} & {70.07} \\
\midrule
DynamiCrafter~\cite{xing2023dynamicrafter}       &  {52.09}  & {47.19} & {25.15} & {47.36} & {25.00} & {72.90} & {60.95} & \ul{78.85} & {54.40} & {41.11} & {39.25} & {26.92} \\
VideoCrafter1-T2V~\cite{chen2023videocrafter1}  &  {47.10}   & {43.54} & {21.61} & {50.44} & {60.78} & {64.86} & {51.36} & {38.05} & {42.63} & {11.76} & \textbf{75.00} & {18.87} \\
VideoCrafter1-I2V~\cite{chen2023videocrafter1}   &  {50.47}  & {47.64} & {25.46} & {24.25} & {35.27} & {74.42} & {73.89} & {65.17} & {54.85} & {55.63}  & {25.00} & {42.49} \\
VideoCrafter2~\cite{chen2023videocrafter1}        &  {52.57} & {47.49} & {28.92} & {39.07} & {72.46} & {65.14} & {61.85} & {43.79} & {56.74} & {47.12}  & {30.40} & {29.39} \\
T2V-Turbo~\cite{li2024t2v}       &   {45.65}     & {40.20} & {27.80}  & {30.68} & {69.14} & {38.72} & {34.84} & {49.65} & \textbf{68.74} & {34.87} & {40.09} & {7.48} \\
EasyAnimate~\cite{xu2024easyanimate}          & {52.85}  & {51.65} & {26.72} & {54.50} & {50.76} & {67.29} & {47.35} & {73.05} & {50.31} & \ul{75.00} & {31.16} & {40.32} \\
Allegro~\cite{zhou2024allegro}               & {55.31} & {51.97} & {24.84} & {57.47} & {51.48} & {70.50} & {69.89} & {65.60} & {47.41} & {54.39} & {40.28} & {37.81} \\
Vchitect-2.0~\cite{fan2025vchitect}          & {42.28} & {38.47} & {26.55} & {49.54} & {65.75} & {41.53} & {42.30} & {25.69} & {44.58} & {33.59} & {33.81} & {21.31} \\
LTX-Video~\citep{hacohen2024ltx} & {55.44} & {56.54} & {25.06} & {53.41} & {39.73} & {78.41} & {88.92} & {53.50} & {49.08} & \textbf{76.22} & {29.95} & \ul{71.09} \\
CogVideoX-T2V~\cite{yang2024cogvideox}        &  {54.18} & {48.79} & {40.22} & {51.05} & {68.12} & {68.81} & {64.20} & {42.19} & {44.67} & {25.00}  & \ul{47.31} & {36.28}\\
CogVideoX-I2V~\cite{yang2024cogvideox}       &  {62.15}  & \textbf{59.12} & {38.27} & {40.07} & {36.73} & {86.21} & {88.12} & \textbf{83.22} & {62.44} & {69.56} & {26.42} & {60.15} \\
\midrule
SceneScape~\cite{fridman2024scenescape}       &    {50.73}   & {35.51} & {84.99}  & {47.44} & {28.64} & {76.54} & {62.88} & {21.85} & {32.75} & {0.00} & {0.00} & {0.00} \\
Text2Room~\cite{hollein2023text2room}         &   {62.10}   & {43.47} & \textbf{94.01} & {38.93} & {50.79} & \ul{88.71} & {88.36} & {37.23} & {36.69} & {0.00} & {0.00} & {0.00} \\
LucidDreamer~\cite{chung2023luciddreamer}       &  \ul{70.40}   & {49.28} & {88.93}  & {41.18} & \textbf{75.00} & \textbf{90.37} & \textbf{90.20} & {48.10} & {58.99} & {0.00} & {0.00} & {0.00} \\
WonderJourney~\cite{yu2024wonderjourney}       & {63.75}   & {44.63} & {84.60}  & {37.10} & {35.54} & {80.60} & {79.03} & {62.82} & \ul{66.56} & {0.00} & {0.00} & {0.00} \\
InvisibleStitch~\cite{engstler2024invisible}   &   {61.12}   & {42.78} & \ul{93.20}  & {36.51} & {29.53} & {88.51} & \ul{89.19} & {32.37} & {58.50} & {0.00} & {0.00} & {0.00} \\
WonderWorld~\cite{yu2024wonderworld}          & \textbf{72.69}  & {50.88} & {92.98}  & {51.76} & {71.25} & {86.87} & {85.56} & {70.57} & {49.81} & {0.00} & {0.00} & {0.00} \\
\midrule
4D-fy~\cite{bahmani20244dfy}                 & {27.98} & {32.10} & {69.92} & {55.09} & {0.85} & {35.47} & {1.59} & {32.04} & {0.89} & {22.22} & {22.88}  & \textbf{80.06}\\
\bottomrule
\end{tabular}%
}
\vspace{-0.1cm}
\caption{\textbf{WorldScore evaluation of 20 world generation models.} Top: Close-source video models. Middle: Open-source video models. Bottom two rows: 3D and 4D models. Abbreviations: Ctrl=Controllability, Align=Alignment, Consist=Consistency, Photo=Photometric, Qual=Quality, Acc=Accuracy, Mag=Magnitude, Smooth=Smoothness.}
\vspace{-0.2cm}
\label{tab:evaluation_results}
\end{table*}

%% file: text/4_exp.tex
\section{Results}

\myparagraph{Validation.} We validate our metrics by human study. Our results suggest that \modelfull's metrics align with human preference, and \modelfull is robust to different video resolutions and aspect ratios. We leave details in Supp.~\ref{supp_sec:metric_eval}.

\myparagraph{Models.}
We evaluate 20 available world generation models on our \modelfull benchmark. We assess 13 video generation models, including two leading commercial closed-source I2V models—Gen-3~\cite{gen3} and Hailuo~\cite{minimax}, along with 7 well-known open-source I2V models: DynamiCrafter~\cite{xing2023dynamicrafter}, VideoCrafter1-I2V~\cite{chen2023videocrafter1}, VideoCrafter2~\cite{chen2024videocrafter2}, EasyAnimate~\cite{xu2024easyanimate}, CogVideoX-I2V~\cite{yang2024cogvideox}, LTX-Video~\citep{hacohen2024ltx} and Allegro~\cite{zhou2024allegro}, and 4 open-source T2V models: VideoCrafter1-T2V, T2v-Turbo~\cite{li2024t2v}, Vchitect-2.0~\cite{fan2025vchitect}, and CogVideoX-T2V. Additionally, we evaluate six well-known 3D scene generation models: SceneScape~\cite{fridman2024scenescape}, Text2Room~\cite{hollein2023text2room}, LucidDreamer~\cite{chung2023luciddreamer}, WonderJourney~\cite{yu2024wonderjourney}, InvisibleStitch~\cite{engstler2024invisible}, and WonderWorld~\cite{yu2024wonderworld}. Moreover, we include an open-source 4D generation model, 4D-fy~\cite{bahmani20244dfy}. We leave details of these models in Table~\ref{supp:world_generation_models} in supplementary material.

\input{fig_text/worldscore_subdomain}

\subsection{Observations and Challenges}

We show the WorldScore benchmark results in Table~\ref{tab:evaluation_results}. We identify key challenges in world generation:

\myparagraph{3D models excel in static world generation.} From the \wss results, we observe that 3D scene generation models generally perform better, e.g., WonderWorld~\citep{yu2024wonderworld} (72.69) and LucidDreamer~\citep{chung2023luciddreamer} (70.40) are the top-2, much better than the best video model CogVideoX-I2V~\citep{yang2024cogvideox} (62.15). This is because 3D models inherently have high camera controllability and, thus, better content alignment due to the larger space they can create, as well as high 3D and photometric consistency. However, they do not allow for the generation of dynamic worlds. When extended to 4D for dynamics, 4D-fy~\citep{bahmani20244dfy} does not perform well, likely due to the intrinsic difficulty in 4D scene generation.

\myparagraph{Video models lack camera controllability.} Even CogVideoX-T2V~\citep{yang2024cogvideox}, the best video generation model in camera controllability (40.22), scored much lower than any 3D/4D generation model. This is the main challenge for video generation models to achieve good static world generation. Recent work in injecting camera conditioning~\citep{wang2024motionctrl,he2024cameractrl} might be a promising solution.

\myparagraph{The best open-source video models are as good as closed-source video models.} Comparing CogVideoX-I2V~\citep{yang2024cogvideox}, with Gen-3 and Hailuo~\citep{minimax}, we observe that CogVideoX-I2V scored even higher than both closed-source models in both \wss (62.15) and \wsd (59.12). However, CogVideoX-I2V is not better than them in every aspect. For instance, we observe that CogVideoX-I2V is better at camera controllability yet worse at object controllability and content alignment.

\myparagraph{Trade-offs in motion smoothness and magnitude}. Comparing motion smoothness and motion magnitude metrics for each method, we observe that larger motion often comes at the cost of lower smoothness, revealing current challenge for video models in maintaining both significant motion and natural transitions.

\myparagraph{Larger motion does not necessarily mean more accurate motion placement.} The correlation between the motion magnitude and accuracy is weak. This implies that models that can produce large motion do not guarantee correct motion placement to follow instructions. Instead, they could hallucinate unintended camera motion or irrelevant motion. More robust motion modeling may be needed to balance the three dynamics metrics.

\myparagraph{Video models are weak in long sequence generation and in outdoor scenes.} We further evaluate model performance across different subdomains, and we show \wss results in Figure~\ref{fig:worldscore_subdomain}. We observe that video generation models struggle significantly with long-sequence (large world generation) tasks. In addition, video models are significantly weaker than 3D models in outdoor scenes, while the gap is smaller in indoor scenes.

\myparagraph{T2V models are easier to steer than I2V models.} Compare T2V models to I2V models, e.g., CogVideoX-T2V and CogVideoX-I2V, we observe that T2V models generally have higher scores in the controllability aspect and larger motion magnitude, while I2V models have higher scores in quality aspect. Through empirical examination, we find that this is because T2V models are willing to generate larger camera motion, while I2V models tend to stick to the input image viewpoint. This reveals a challenging in controlling I2V models to generate new scene contents. We leave further visualizations in Supp.~\ref{supp_sec:more_results}.

%% file: fig_text/worldscore_subdomain.tex
\begin{figure}
    \centering
    \includegraphics[width=0.47\textwidth]{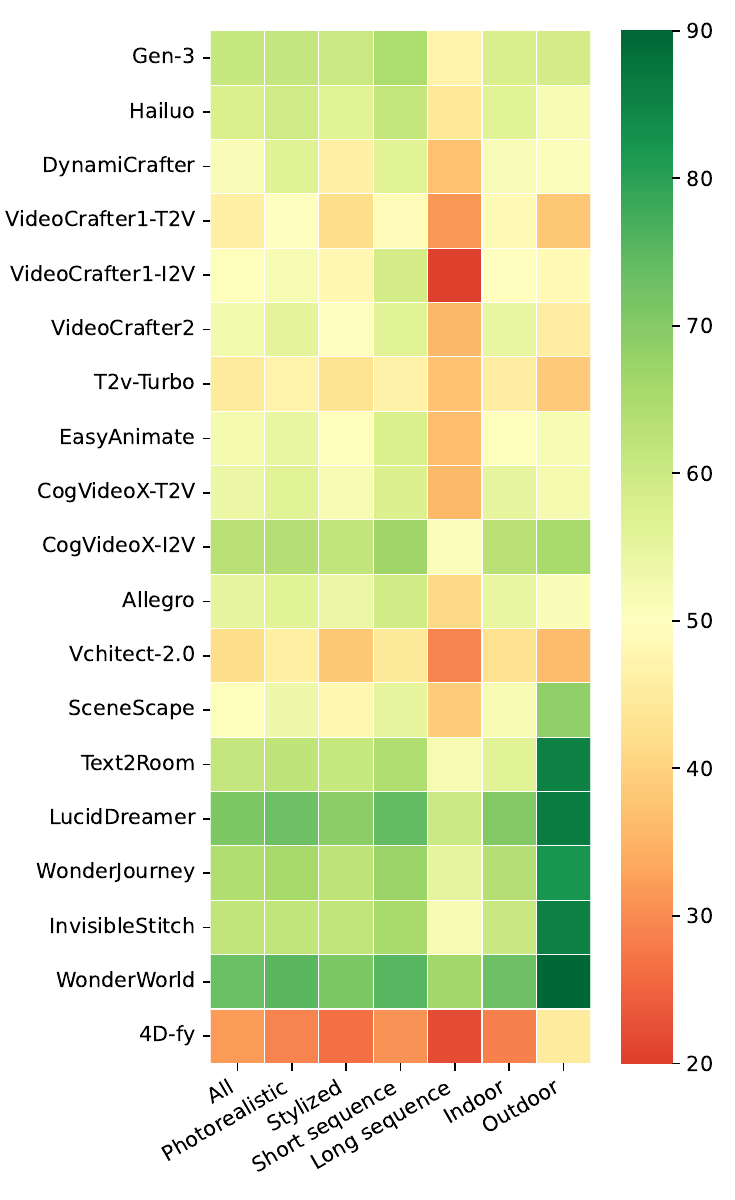}
    \vspace{-0.2cm}
    \caption{\wss across different subdomains.} 
    \vspace{-0.2cm}
    \label{fig:worldscore_subdomain}
\end{figure}

%% file: text/5_conclusion.tex
\section{Conclusion}
The WorldScore benchmark reveals current limitations in world generation approaches. For 3D models, while they excel in static world generation, extending them to 4D representations and incorporating dynamics remains challenging. For video models, the main challenges include controllability, long-sequence generation, and generating outdoor scenes. These insights point to directions for future research: bridging the gap between 3D and 4D representations, developing more robust controllability mechanisms, and designing architectures capable of handling extended scene sequences. We believe the WorldScore benchmark will serve as a valuable tool for measuring progress toward more capable and versatile world generation systems.

%% file: supp.tex
\clearpage
\renewcommand\thefigure{S\arabic{figure}}
\setcounter{figure}{0}
\renewcommand\thetable{S\arabic{table}}
\setcounter{table}{0}
\renewcommand\theequation{S\arabic{equation}}
\setcounter{equation}{0}
\pagenumbering{arabic}%
\renewcommand*{\thepage}{S\arabic{page}}
\setcounter{footnote}{0}
\setcounter{page}{1}
\maketitlesupplementary
\appendix

\input{supp_text/1_world_specification}

\input{supp_text/2_dataset_construction}

\input{supp_text/3_metrics}
\input{supp_text/5_validation}

\input{supp_text/4_further_viz}

%% file: supp_text/1_world_specification.tex
\section{Additional Details on World Specification}
\label{supp_sec:supplementary_world_specification}

\input{fig_text/supp_model_family}

We provide additional details on world specification pre-processing $w_\text{proc}$ in Eq.~\ref{eq:world_generation}. We evaluate models across 3D scene generation, 4D scene generation, and video generation, each with distinct input requirements. For instance, 3D/4D scene generation models~\cite{yu2024wonderjourney, yu2024wonderworld} accept precise camera poses as input, whereas video generation models do not. Also, among these models, some are T2V models~\cite{li2024t2v, fan2025vchitect}, which rely solely on text-based control, while others are I2V models~\cite{yu2024wonderworld, xu2024easyanimate, yu2024wonderjourney, gen3, minimax}, which accept image control signals. To accommodate these variations, $w_\text{proc}$ ensures that each model receives inputs in its appropriate format.

Specifically, $w_\text{proc}$ standardizes the inputs as follows: 

\begin{itemize}
    \item \textbf{Reference image $\mathbf{I}$:} The image for current scene $\mathcal{C}$ is center-cropped and resized to match the resolution required by each model (see Table~\ref{supp:world_generation_models} for the specific resolutions). This serves as both a visual style reference and a necessary input for I2V models. Notably, T2V models are treated as I2V models that ignore image-based control signals.

    \item \textbf{Layout $\mathcal{L}$:} The world specification module generates a predefined precise camera trajectory $\mathcal{T}$ (which serve as ground truth for camera controllability) and corresponding textual descriptions $\mathcal{Y}$ (\eg, \textit{``camera moves left''}) as world layout $\mathcal{L}$. $w_\text{proc}$ gives models that accept explicit camera control signals the transformed camera poses $\mathcal{T}^{\prime}$, ensuring alignment across different camera types, while models without explicit camera control receive textual descriptions $\mathcal{Y}$ instead. 

    \item \textbf{Next-scene prompt $\mathcal{N}$:} For 3D/4D models which all accept camera matrices as input, $w_\text{proc}$ does not adapt the prompt $\mathcal{N}$. For video models that do not accept camera matrices as input, $w_\text{proc}$ processes the next-scene prompt $\mathcal{N}$ by adding camera movement text to it.
\end{itemize}

%% file: fig_text/supp_model_family.tex
\begin{table*}[ht]
\centering
\small
\setlength{\tabcolsep}{4pt}
\renewcommand{\arraystretch}{1.2} 
\begin{tabular}{l c c c c c c c c}
\toprule
{Method} & {Version} & {Ability} & {Resolution} & {Length (s)} & {FPS} & {Open Source} & {Speed$^\dagger$} & {Camera$^\S$} \\
\midrule
Gen-3 \cite{gen3} & 24.07.01 & I2V & 1280×768 & 10 & 24 & \xmark & 1 min & \xmark \\
Hailuo \cite{minimax} & 24.08.31 & I2V & 1072×720 & 5.6 & 25 & \xmark & 3.5 min & \xmark \\
\midrule
DynamiCrafter \cite{xing2023dynamicrafter} & 23.10.18 & I2V & 1024×576 & 5 & 10 & \cmark & 2.5 min & \xmark \\
\multirow{2}{*}{VideoCrafter1 \cite{chen2023videocrafter1}} & \multirow{2}{*}{23.10.30} & T2V & 1024×576 & 2 & 8 & \cmark & 7 min & \xmark \\
 &  & I2V & 512×320 & 2 & 8 & \cmark & 2 min & \xmark \\
VideoCrafter2 \cite{chen2024videocrafter2} & 24.01.17 & T2V & 512×320 & 2 & 8 & \cmark & 2 min & \xmark \\
T2V-Turbo \cite{li2024t2v} & 24.05.29 & T2V & 512×320 & 3 & 16 & \cmark & 5 s & \xmark \\
EasyAnimate \cite{xu2024easyanimate} & 24.05.29 & I2V & 1344×768 & 6 & 8 & \cmark & 16 min & \xmark \\
\multirow{2}{*}{CogVideoX \cite{yang2024cogvideox}} & \multirow{2}{*}{24.08.12} & T2V & 720×480 & 6 & 8 & \cmark & 2.4 min & \xmark \\
 &  & I2V & 720×480 & 6 & 8 & \cmark & 2.4 min & \xmark \\
Allegro \cite{zhou2024allegro} & 24.10.20 & I2V & 1280×720 & 6 & 15 & \cmark & 0.5 h & \xmark \\
Vchitect-2.0 \cite{zhou2024allegro} & 25.01.14 & T2V & 768×432 & 5 & 8 & \cmark & 2.8 min & \xmark \\
LTX-Video \cite{hacohen2024ltx} & 25.05.05 & I2V & 768×512 & 4 & 30 & \cmark & 2.4 min & \xmark \\
\midrule
SceneScape \cite{fridman2024scenescape} & 23.02.02 & T2V & 512×512 & 5 & 10 & \cmark & 11.4 min & \cmark \\
Text2room \cite{hollein2023text2room} & 23.03.21 & I2V & 512×512 & 5 & 10 & \cmark & 12.4 min & \cmark \\
LucidDreamer \cite{chung2023luciddreamer} & 23.11.22 & I2V & 512×512 & 5 & 10 & \cmark & 6.4 min & \cmark \\
WonderJourney \cite{yu2024wonderjourney} & 23.12.06 & I2V & 512×512 & 5 & 10 & \cmark & 6.3 min & \cmark \\
InvisibleStitch \cite{engstler2024invisible} & 24.04.30 & I2V & 512×512 & 5 & 10 & \cmark & 2.3 min & \cmark \\
WonderWorld \cite{yu2024wonderworld} & 24.06.13 & I2V & 512×512 & 5 & 10 & \cmark & 10 s & \cmark \\
\midrule
4D-fy \cite{bahmani20244dfy}$^*$ & 23.11.29 & T2V & 256×256 & 4 & 30 & \cmark & 3 h & \cmark \\
\bottomrule
\end{tabular}
\caption{\textbf{Further details of the world generation models in our benchmark.} $^\dagger$ The reported values indicate the average generation time per instance. All generations were conducted on H100 and L40S GPUs. $^\S$ This indicates whether the model accepts precise camera poses as input. $^*$ For 4D-fy, it takes about 20 hours for each generation, so we decrease the iteration steps to save time. While these models use different output resolutions and aspect ratios, our validation shows that \modelfull metrics are robust against these differences (Sec.~\ref{supp_sec:metric_eval}).}
\label{supp:world_generation_models}
\end{table*}

%% file: supp_text/2_dataset_construction.tex
\section{Additional Details on Dataset Curation}
\label{supp_dataset}

\input{fig_text/supp_existing_dataset}
\input{fig_text/supp_filtration}

\subsection{Image Filtering}\label{sec:filtering}

\input{fig_text/supp_stylized_images}

To construct a high-quality and diverse image dataset as our starting current scene images, we source from both existing datasets and supplement them with Unsplash~\citep{unsplash}.
Existing scene datasets~\cite{song2015sun, Matterport3D, vasiljevic2019diode, eth3d, lhq,zhu2022learninginterviorverse, roberts2021hypersim, le2021eden} (Table~\ref{supp_tab:supp_datasets}) are designed for scene understanding~\cite{song2015sun, Matterport3D, vasiljevic2019diode, roberts2021hypersim}. Many of the images in these datasets are not suitable as the current scene image, as they may contain excessive redundancy, unusual viewpoints, and narrow-angle perspectives. Therefore, we apply filtering based on several criteria (see Figure~\ref{supp_fig:filtration} for visualization of the filtering):

\paragraph{Quality.} We employ CLIP-IQA~\cite{wang2023exploringclipiqa} and CLIP Aesthetic~\cite{schuhmann2022clip+aesthetic} predictors to filter out images with poor visual quality.
    
\paragraph{Perspective.} To ensure appropriate viewpoint composition, we utilize the Perspective Fields~\cite{jin2023perspectivefields} to model the local perspective properties (\eg, yaw, pitch, and FOV). We filter out images with extreme roll or pitch angles and those with a narrow FOV, aiming to retain open-angle, front-facing perspectives.
    
\paragraph{Similarity.} Since many datasets contain redundant sequential images, we use CLIPSIM \cite{radford2021learningclipsim} to remove visually similar images.

\paragraph{Brightness.} To exclude overly dark images, we compute image brightness and filter out those below a predefined threshold.
    
\paragraph{Human Judgment.} Finally, we conduct a manual review to refine the selection, ensuring the curated images align with human perception and the intended use case.

\subsection{Stylized Image Generation}\label{sec:stylized}

After filtering and categorization, we obtain our photorealistic image dataset. Then, for each photorealistic image, we generate a stylized counterpart image using a text-to-image model~\cite{recraft}. 

\paragraph{Predefined style sets.} To ensure diversity of visual style, we curate a  predefined style set by referencing visual art history \cite{fineart}, supplemented with commonly used visual styles from SDXL \cite{sdxl}. Our final selection includes: \textit{anime}, \textit{cyberpunk}, \textit{Chinese ink painting}, \textit{ukiyo-e}, \textit{impressionism}, \textit{post-impressionism}, and \textit{minecraft}. See example images in Figure~\ref{supp_fig:stylized_images}.

\input{fig_text/supp_llm_query_output}

\subsection{Next-Scene Text Prompts Curation}

\label{supp_sec:prompt_suit_details}
We use GPT-4o \cite{gpt4o} for scene description generation, with distinct approaches for static and dynamic scenarios. Specifically, for the static world generation task, we employ an auto-regressive process using the following task specification $\mathcal{J_{\text{static}}}$ for system calls:

\textit{``You are an intelligent scene generator. Imaging you are wondering through a sequence of scenes, please tell me what sequentially next scene would you likely to see? You need to generate 1 to 3 most prominent entities in the scene. The scenes are sequentially interconnected, and the entities within the scenes are adapted to match and fit with the scenes. You also have to generate a brief scene description. If needed, you can make reasonable guesses. Please ensure the output is in the following JSON format: \{`Entities': [`entity\_1', ...], `Prompt': `scene description'\}.''}

For the dynamic world generation task, we use the task specification $\mathcal{J_{\text{dynamic}}}$ for single system call:

\textit{``You are an intelligent motion dreamer, capable of identifying the objects within an image that can exhibit dynamic motion. I will provide you with an image, and your task is to identify the most prominent object(s) that have the potential for dynamic movement. You also have to briefly describe how the object(s) move. If needed, you can make reasonable guesses. Please ensure the output is in the following JSON format: \{'Objects': ['object\_1', ...], 'Prompt': 'description of how the object(s) move'\}.''} 

We show an example of generated next-scene prompts in Table~\ref{supp_tab:query_outputs_examples}.

%% file: fig_text/supp_existing_dataset.tex
\begin{table}[t]
\resizebox{0.5\textwidth}{!}{%
\centering
\small
\begin{tabular}{clccc}
\toprule
\rotcell{Scene\\Type} & \rotcell{Dataset} & \rotcell{Image\\Type} & \rotcell{Res.} & \rotcell{\# Images} \\
\midrule
\multirow{5}{*}{Indoor} 
    & InterviorVerse \cite{zhu2022learninginterviorverse} & Synthetic & 640×480 & 50,000 \\
    & Hypersim \cite{roberts2021hypersim}      & Synthetic & 1024×768  & 77,400 \\
    & SUN-RGBD \cite{song2015sun}      & Real      & 640×480  & 10,000 \\
    & Matterport3D \cite{Matterport3D}  & Real      & 1280×1024 & 194,400 \\
    & DIODE-indoor \cite{vasiljevic2019diode}         & Real      & 1024×768   & 9,052 \\
    & ETH3D-indoor \cite{eth3d}         & Real      & 6214×4138 & 597 \\
\midrule
\multirow{5}{*}{Outdoor} 
    & LHQ \cite{lhq}            & Real      & 1024×1024 & 90,000 \\
    & EDEN \cite{le2021eden}          & Synthetic & 640×480 & 300,000 \\
    & Argoverse-HD \cite{argoverse}   & Real      & 1920×1200  & 70,000 \\
    & DIODE-outdoor \cite{vasiljevic2019diode}         & Real      & 1024×768   & 18,206 \\
    & ETH3D-outdoor \cite{eth3d}         & Real      & 6214×4138 & 301 \\
\bottomrule
\end{tabular}
}
\caption{{Statistics of the scene datasets we source from.}}
\label{supp_tab:supp_datasets}
\end{table}

%% file: fig_text/supp_filtration.tex
\begin{figure*}[t]
    \centering
    \includegraphics[width=1\textwidth]{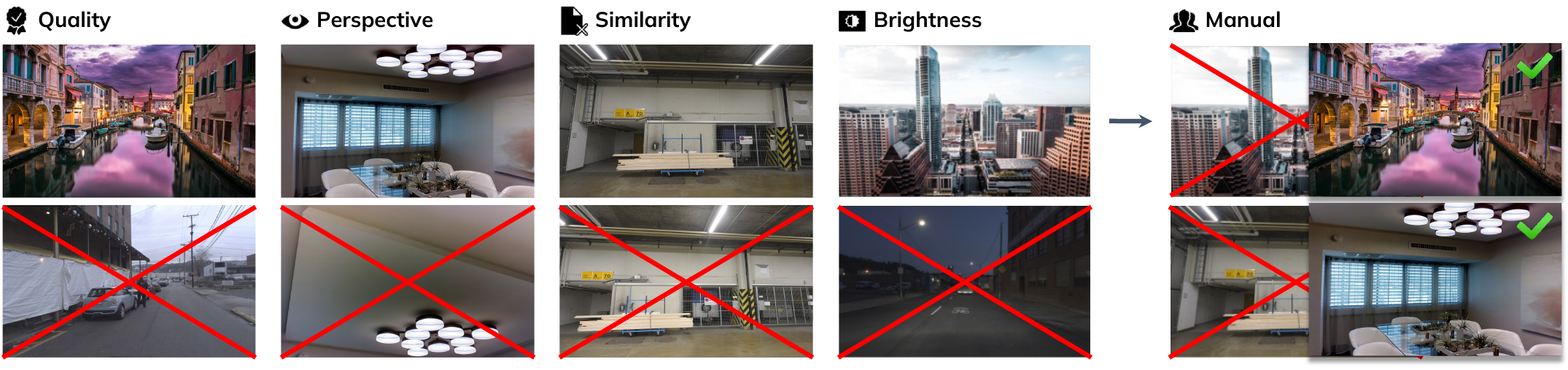}
    \caption{\textbf{Filtering.} We apply the filtering based on several criteria to remove undesired images. Besides automatic metrics, we also apply a final manual inspection to remove infeasible world generation starting scenes such as the mid-air city image in the 4th column.}
    \label{supp_fig:filtration}
\end{figure*}

%% file: fig_text/supp_stylized_images.tex
\begin{figure*}[ht]
    \centering
    \includegraphics[width=0.9\textwidth]{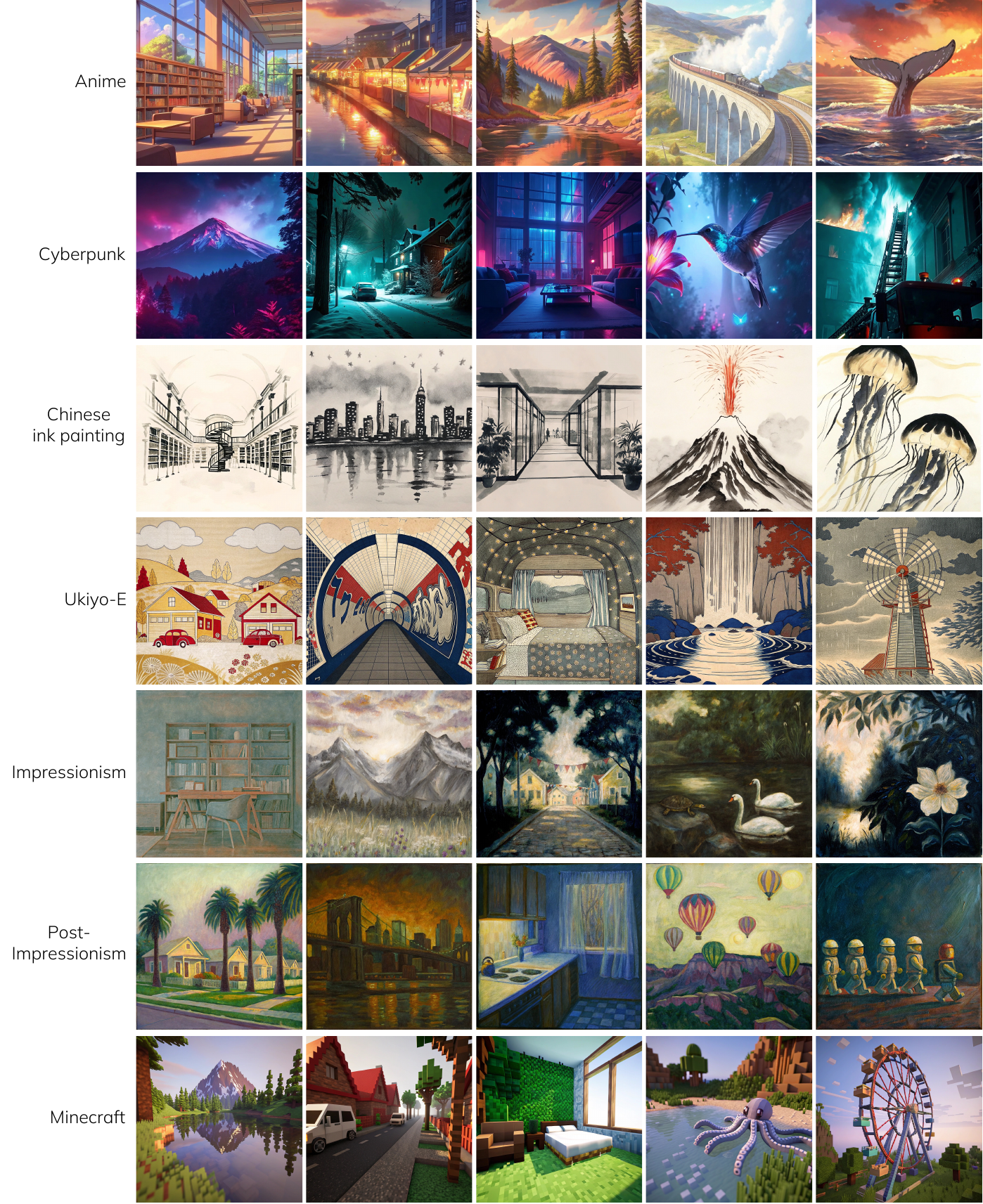}
    \caption{\textbf{Examples of stylized images.} Our predefined style set contain 7 different visual art styles.}
    \label{supp_fig:stylized_images}
\end{figure*}

%% file: fig_text/supp_llm_query_output.tex
\begin{table}[ht]
    \centering
    \begin{tcolorbox}[colframe=gray!50!white, colback=white!5!white, title=\textbf{Generated Next-Scene Prompt $\mathcal{N}$}]
        
        \textbf{Static world generation}  
        
        \textbf{\# Scene 1} \\
        \textit{\{``Entities'': [``yellow armchair'', ``bookshelf''], ``Prompt'': ``A Cozy Reading Nook with a plush Yellow Armchair surrounded by a towering Bookshelf filled with books.''\}}

        \textbf{\# Scene 2} \\
        \textit{\{``Entities'': [``potted plants''], ``Prompt'': ``A serene Tranquil Garden Patio featuring a cozy Yellow Armchair surrounded by lush Potted Plants gently swaying in a soft, breezy atmosphere.''\}}

        \textbf{\# Scene 3} \\
        \textit{\{``Entities'': [``wooden rail''], ``Prompt'': ``A Rustic Balcony Retreat featuring a cozy Yellow Armchair and a classic Wooden Rail bathed in the warm glow of the setting sun.''\}}

        \tcblower
        
        \textbf{Dynamic world generation}  
        
        \textit{\{``Objects'': [``windmill'', ``cloud'', ``sea''], ``Prompt'': ``The windmill blades spin in a circular motion driven by the wind, creating a consistent rotational movement. The clouds drift slowly across the sky, pushed gently by the breeze. The sea surface ripples and undulates, as small waves ripple across its surface.''\}}

    \end{tcolorbox}
    \caption{An example of generated next-scene prompt for static and dynamic world generation. The ``prompt'' in the above box is the next-scene prompt $\mathcal{N}$. The ``entities'' are the objects to detect when computing object controllability. The ``objects'' are used to help annotate the motion masks for computing motion accuracy.} 
    \label{supp_tab:query_outputs_examples}
\end{table}

%% file: supp_text/3_metrics.tex
\input{fig_text/dataset_table}

\section{Additional Details on Metrics}
\label{supp_sec:supplementary_of_worldscore_eval}

\subsection{Camera Controllability}
\label{supp_sec:camera_control}

As formulated in Eq.~\ref{algorithm:camera_error_geometric_mean}, we combine $e_{\theta}$ and $e_t$ with geometric mean to calculate the camera error. Specifically, we estimate the frame-wise camera poses using DROID-SLAM \cite{teed2021droid}. Then we compute the angular deviation between the ground truth and the estimated camera rotations (in degrees):
\begin{equation}
    e_{\theta} = \arccos\left(\frac{\text{tr}(\mathbf{R}_{\text{gt}}\mathbf{R}^T) - 1}{2}\right) \cdot \frac{180}{\pi},
    \label{supp_algorithm:rotation_errors}
\end{equation}
and the scale-invariant Euclidean distance between ground truth and estimated camera positions:
\begin{equation}
    e_t = \|\mathbf{t}_{\text{gt}} - s\mathbf{t}\|_2,
    \label{supp_algorithm:translation_errors}
\end{equation}
where $\mathbf{R}_{\text{gt}}, \mathbf{R} \in SO(3)$ denote the ground truth and estimated rotation matrices, $\mathbf{t}_{\text{gt}}, \mathbf{t} \in \mathbb{R}^3$ denote the ground truth and estimated camera positions, and $s$ denotes the least-square scale.

The final camera controllability error for a model is computed by averaging the error $e_\text{camera}$ over all frames of all generated videos.

\subsection{3D Consistency}
\label{supp_sec:3d_consistency}

To quantify the 3D consistency of generated videos, we use DROID-SLAM \cite{teed2021droid} to do the reconstruction and calculate the reprojection error. One key advantage of DROID-SLAM is its dense nature. Unlike sparse methods such as COLMAP \cite{colmap1, colmap2}, which rely on selecting ``good'' feature matches while discarding the rest, DROID-SLAM employs a differentiable Dense Bundle Adjustment (DBA) layer. This layer continuously refines camera poses and dense, per-pixel depth estimates to ensure consistency with the current optical flow. By leveraging all available points, rather than focusing on partial matches, this dense approach aligns with our goal of assessing 3D consistency across the entire scene. This evaluation dimension ensures a more comprehensive understanding of the spatial coherence in generated videos.

Specifically, we calculate the reprojection error after DBA layer refinement:
\begin{equation}
    e_{\text{reproj}} = \frac{1}{|\mathcal{V}|}\sum_{(i,j) \in \mathcal{V}} \left\| \mathbf{p}^*_{ij} - \Pi (\mathbf{P}_{ij}) \right\|_2,
    \label{supp:reprojection_error}
\end{equation}
where $\mathcal{V}$ denotes the valid set of co-visible points, $\mathbf{p}^*_{ij}$ is the observed point on the ground truth image, $\mathbf{P}_{ij}$ is the reconstructed 3D point, obtained from refined depth and camera pose, $\left\| \cdot \right\|_2$ calculates the Euclidean distance.

\subsection{Photometric Consistency}
\label{supp_sec:photometric_consistency}
The photometric consistency metric is to quantify the model capability to generate stable visual appearances.
We estimate the optical flow between consecutive frames and compute the Average End-Point Error (AEPE). Specifically, given two consecutive frames $A$ and $B$, we first track a set of center-cropped points $\mathbf{p}_A$ from frame $A$ to frame $B$ using forward optical flow $\mathcal{F}_{A\to B}$:
\begin{equation} 
\mathbf{p}_B = \mathbf{p}_A + \mathcal{F}_{A\to B}(\mathbf{p}_A). 
\end{equation}

We then track the same points back from frame $B$ to frame $A$ using backward optical flow $\mathcal{F}_{B\to A}$:

\begin{equation} 
\mathbf{p}_{A}^{\prime} = \mathbf{p}_B + \mathcal{F}_{B\to A}(\mathbf{p}_B). 
\end{equation}

Ideally, if the object remains photometrically consistent, the tracked points should return to their original locations, \ie, $\mathbf{p}_{A}^{\prime} \approx \mathbf{p}_{A}$. we quantify the deviation using the AEPE:
\begin{equation} 
e_\text{photometric} = \frac{1}{N} \sum_{i=1}^{N} \left\| \mathbf{p}_{A,i} - \mathbf{p}_{A,i}^{\prime} \right\|_2, 
\end{equation}
where $N$ is the number of sampled points. A higher AEPE indicates greater photometric inconsistency, signaling anomalies such as identity shifts, texture flickering, or object disappearances. 
Finally, the photometric consistency error is computed by averaging $e_\text{photometric}$ over all consecutive frame pairs of all generated videos.

\subsection{Subjective Quality}
\label{supp_sec:subjective_quality}

Numerous trained image quality assessment metrics exist, such as CLIP-Aesthetic \cite{schuhmann2022clip+aesthetic} and QAlign-Aesthetic \cite{qalign}, which focus on factors like layout composition, color harmony, realism, and artistic appeal. Additionally, image quality predictors like MUSIQ \cite{ke2021musiq} and CLIP-IQA \cite{wang2023exploringclipiqa} evaluate distortions such as overexposure, noise, and blur.

Our goal is to use automatic metrics that align well with human perception to evaluate the subjective quality of generated scenes. To identify the (combination of) best subjective quality predictors, we systematically conduct a human preference study the pick the one that best matches human perception on world generation quality. We find that the combination (arithmetic mean) of CLIP-IQA+ \cite{wang2023exploringclipiqa} and CLIP Aesthetic \cite{schuhmann2022clip+aesthetic} works the best. We show more details in Sec.~\ref{supp_sec:metric_eval}.

\subsection{Motion Accuracy} 
We assess whether motion occurs in the intended regions by:
\begin{equation} 
    s_\text{motion-acc} = \max \left( \mathbf{F} \odot \mathbf{M} \right) - \max \left( \mathbf{F} \odot \bar{\mathbf{M}} \right),
    \label{eq:motion_accuracy} 
\end{equation}
where $\mathbf{F}\in\mathbb{R}^{H\times W}$ denotes the magnitude of optical flow between a pair of consecutive frames in the generated video $\mathbf{V}$ estimated by SEA-RAFT \cite{wang2024searaft}, $\mathbf{M}\in\{0,1\}^{H\times W}$ denotes the segmentation masks at the former frame which has $1$ at the pixels of dynamic objects, and the $\max$ operator picks the maximum value among all the entries of a matrix. We track the mask of dynamic objects $\mathbf{M}$ using SAM2 \cite{ravi2024sam2}, where the first-frame segmentation masks are provided in our dataset. The final motion accuracy score is computed by averaging $s_\text{motion-acc}$ across all pairs of consecutive frames of all generated videos.

\subsection{Motion Magnitude}

Some models take a ``conservative'' approach, generating only subtle motion. While the output appears visually smooth and high-quality, the motion is often minimal and uninteresting. Some models even produce near-static videos despite prompts explicitly describing motion. We measure this with $s_\text{motion-mag}$, defined as the median value of all the entries of $\mathbf{F}$, and the final motion magnitude metric is the average of $s_\text{motion-mag}$ across all pairs of consecutive frames of all generated videos.

\subsection{Motion Smoothness}
\label{supp_sec:motion_smoothness}

We leverage the motion priors from a standard video frame interpolation models \cite{zhang2024vfimamba} to evaluate the smoothness of generated motion. Specifically, given a generated video consisting of frames $\{\mathbf{f}_0,\mathbf{f}_1,\mathbf{f}_2,\cdots\}$, we drop the odd-indexed frames $\{\mathbf{f}_1,\mathbf{f}_3,\cdots\}$ to obtain a lower frame rate video, and then we use video frame interpolation to infer the dropped frames. Finally, we compute the mean squared error, SSIM \cite{wang2004imagessim}, and LPIPS \cite{zhang2018unreasonablelpips} between the reconstructed frames and the original dropped frames. After each metric score is computed and normalized (Supp.~\ref{supp_sec:score_normalization_mapping}), we average them to get the motion smoothness metric.

\subsection{Empirical Bounds}
\label{supp_sec:empirical_bounds}

In this section, we discuss how we calculate the empirical bounds for each evaluation dimension, which will be used for linear normalization in Supp.~\ref{supp_sec:score_normalization_mapping}.

\paragraph{Empirical bounds for camera controllability.} Since the camera controllability metric calculates the deviation between the ground truth and estimated camera poses, the empirical minimum is naturally $0$, which also represents the theoretical lower bound. To approximate the highest achievable values, we use a sequence of fixed cameras as a baseline. This effectively penalizes poorly performing world generation that fails to exhibit any camera movement.

\paragraph{Empirical bounds for object controllability.} Since we evaluate object controllability using the object detection rate, the empirical minimum and maximum are naturally $0$ and $100\%$, respectively, which also represent the theoretical bounds.

\paragraph{Empirical bounds for 3D consistency, style consistency, and photometric consistency.} To establish empirical bounds for these frame-wise metrics, we randomly sample image pairs from our dataset and generate videos by interpolating intermediate frames using a video frame interpolation model \cite{zhang2024vfimamba}. This serves as a baseline exhibiting significant style shifts, low 3D consistency, and poor photometric stability. We define this baseline as empirical maximum for all three metrics, while the empirical minimum for each is set to 0, which is also theoretical minimum.

\paragraph{Empirical bounds for motion smoothness.} To determine empirical values for \textit{motion smoothness}, we leverage high-quality real-world videos. Given that most world generation models produce 3-10 second videos, we retrieve comparable video clips from OpenVid-1M \cite{nan2024openvid}, a large-scale, high-quality video dataset. Specifically, for each prompt in our benchmark, we retrieve the top five OpenVid-1M videos with the highest semantic similarity using CLIP-based text feature matching. Only 3-10 second clips are considered to ensure consistency with the length of generated videos.

Then, we use the retrieved videos as a reference. We manually drop the odd frames and apply bilinear interpolation to reconstruct them. This serves as a baseline, where the resulting interpolated videos represent the ``empirical worst'' (empirical maximum for MSE and LPIPS and empirical minimum for SSIM). The ``empirical best'' is set to 0, indicating perfectly smooth motion.

\paragraph{Empirical bounds for content alignment, subjective quality, motion accuracy, and motion magnitude.} For these four metrics, defining appropriate empirical bounds is challenging. To address this, we apply z-score rescaling, setting the empirical best and worst values so that the performance of selected models falls within the 25 to 75 range. This approach enhances differentiation and ensures a more reliable evaluation.

\subsection{Score Normalization and Mapping}
\label{supp_sec:score_normalization_mapping}

The detailed formulation for score normalization and mapping is as follows:

\begin{equation}
s^{\text{norm}} =
\begin{cases}
\left\langle \frac{s - b^\text{min}}{b^\text{max} - b^\text{min}} \right\rangle, & \text{if higher better}, \\
\left\langle 1 - \frac{s - b^\text{min}}{b^\text{max} - b^\text{min}} \right\rangle, & \text{if lower better},
\end{cases}
\label{eq:normalization}
\end{equation}
where \(s\) denotes the raw value of a given metric, \(b^\text{min}\) and \(b^\text{max}\) denote the empirical bounds of the metric, and $\left\langle \cdot \right\rangle$ denotes the clip function, making sure the normalized score \(s^{\text{norm}}\) is within the range \([0, 1]\), where a higher value corresponds to better performance.

%% file: fig_text/dataset_table.tex
\begin{table*}[ht]
\centering
\small
\setlength{\tabcolsep}{8pt} 
\renewcommand{\arraystretch}{1.3} 
\begin{tabular}{c llcc}
\toprule
\multirow{5}{*}{{Static}}  
    & \textbf{Visual Style} & \textbf{Scene Type} & \textbf{Category} & \textbf{\# Samples} \\ 
\cmidrule(lr){2-5}
    & \multirow{2}{*}{Photorealistic} & Indoor  & Dining, Living, Passage, Public, Work & $5 \times 100$ \\  
    &                                 & Outdoor & City, Suburb, Aquatic, Terrestrial, Verdant & $5 \times 100$ \\  
    \cmidrule(lr){2-5}
    & \multirow{2}{*}{Stylized}       & Indoor  & Dining, Living, Passage, Public, Work & $5 \times 100$ \\  
    &                                 & Outdoor & City, Suburb, Aquatic, Terrestrial, Verdant & $5 \times 100$ \\  
\midrule
\multirow{3}{*}{{Dynamic}}  
    & \textbf{Visual Style} & \multicolumn{2}{c}{\textbf{Motion Type}}  & \textbf{\# Samples} \\  
\cmidrule(lr){2-5}
    & Photorealistic & \multicolumn{2}{l}{Articulated, Deformable, Fluid, Rigid, Multi-Motion} & $5 \times 100$ \\  
    & Stylized       & \multicolumn{2}{l}{Articulated, Deformable, Fluid, Rigid, Multi-Motion} & $5 \times 100$ \\  
\midrule
\multicolumn{4}{r}{\textbf{\# Total Samples}} & {3000} \\  
\bottomrule
\end{tabular}
\caption{\textbf{Dataset Statistics.} We curate a dataset of 3000 test samples that span diverse worlds: static and dynamic, photorealistic and stylized, indoor and outdoor. The static subset is further divided into 5 indoor and outdoor scene categories, while the dynamic subset is categorized by 5 motion types.}
\label{tab:dataset_stats}
\end{table*}

%% file: supp_text/5_validation.tex
\section{Validation with Human Preference}\label{supp_sec:metric_eval}
We validate the \modelfull metrics by human preference study for three purposes: Firstly, we use human preference to select the best combination of subjective quality metrics (e.g., image quality assessment metrics and aesthetic metrics) to form a single ``subjective quality''. Secondly, we use human preference to validate other \modelfull metrics. Lastly, we measure how robust are the metrics to different resolutions and aspect ratios. In particular, we use the following agreement score.

\paragraph{Human preference agreement score.} 
To measure how well each metric aligns with human preferences, we adopted a probabilistic agreement score. Given a video pair $(\text{A}, \text{B})$, a participant is forced to choose one video that appears to have higher subjective quality to them, a.k.a. 2-alternative forced choice (2AFC). We denote the portion of all participants who preferred A as $p$, therefore the portion of all participants who preferred B is $1-p$. Then, consider an automatic assessment metric $m$:
\begin{itemize}
\item If the metric $m$ assigned a higher score to A, i.e., $\text{score}_m(A) > \text{score}_m(B)$, then the agreement score for this pair $(\text{A}, \text{B})$ is $p$.
\item If the metric $m$ assigned a higher score to B, i.e., $\text{score}_m(A) < \text{score}_m(B)$, then the agreement score for this pair $(\text{A}, \text{B})$ is $1-p$.
\item If the metric assigned equal scores to A and B, then the agreement score was set to 0.5.
\end{itemize}%
The final agreement score for each metric was obtained by averaging the agreement scores across all human-rated pairs.

To prepare the pairs of videos for human participants,
we randomly sampled videos generated from CogVideoX-I2V, VideoCrafter1-I2v, DynamiCrafter, WonderJourney, and InvisibleStitch. Each comparison consisted of a pair of videos from different models. We recruited 400 participants for the human study.

Note that in our human preference study, we only use a single question, asking the participant ``which video has higher quality''. While there are possibly different dimensions of subjective quality such as aesthetic quality and perceptual quality, our preliminary human preference study indicates that general human raters often struggle to differentiate between specific dimensions, yielding a very high correlation between aesthetic quality and perceptual quality. Therefore, we only use a single question.

\paragraph{Agreement results on subjective quality.} We show the agreement results in Table~\ref{supp_tab:human_annotation_correlation_value}. Since the combination (arithmetic mean) of CLIP-IQA+ \cite{wang2023exploringclipiqa} and CLIP Aesthetic \cite{schuhmann2022clip+aesthetic} metrics yield the highest agreement, we use this combination to compute our subjective quality.  

\paragraph{Agreement results on other metrics.} To validate other metrics, we divide them into different score buckets, i.e., $90\pm5$, $60\pm5$, and $30\pm5$; and then we compare between buckets. We show results in Table~\ref{tab:re_30_60_90}. The 2AFC results show the our metrics align well with human perception, so that a higher score (both ``90 over 60'' and ``60 over 30'') consistently correlate to a higher human preference. 

\paragraph{Robustness against different resolutions and aspect ratios.} We validate if the metrics are robust to different resolutions and aspect ratios because models vary in these aspects. We use the videos generated by the highest-resolution model (EasyAnimate, $1344\times768$) and apply center-cropping and resizing to a create a version with small resolution ($256\times256$). We evaluate both versions and show results in Table~\ref{tab:re_resolution}. The differences in all metrics are very small ($\leq0.83$), suggesting that our metrics are robust to these differences.

\input{fig_text/supp_table_human_annotation}

\input{fig_text/re_30_60_90}

\input{fig_text/re_resolution}

%% file: fig_text/supp_table_human_annotation.tex
\begin{table}[t]
\centering
\small
\setlength{\tabcolsep}{6pt} %
\renewcommand{\arraystretch}{1.2} %
\begin{tabular}{l c}
\toprule
\textbf{Metric} & \textbf{Correlation} \\ 
\midrule
CLIP-IQA & 0.596 \\
CLIP-IQA+ & 0.602 \\
QAlign Quality & 0.581 \\
QAlign Video Quality & 0.571 \\
MUSIQ & 0.530 \\
CLIP Aesthetic & 0.628 \\
QAlign Aesthetic & 0.479 \\
QAlign Video Aesthetic & 0.556 \\
\midrule
CLIP-IQA+ \& QAlign Quality & 0.582 \\
CLIP Aesthetic \& QAlign Video Aesthetic & 0.629 \\
CLIP-IQA+ \& CLIP Aesthetic & \textbf{0.637} \\
\midrule
{Upper Bound} & 0.772 \\
\bottomrule
\end{tabular}
\caption{\textbf{Agreement of automatic assessment metrics with human preference.} The upper bound is the highest possible agreement score when a metric always agrees with the majority vote for every 2AFC pair.}
\label{supp_tab:human_annotation_correlation_value}
\end{table}

%% file: fig_text/re_30_60_90.tex
\begin{table}[t]
\centering
\scriptsize
\setlength{\tabcolsep}{3pt} 

\begin{tabular}{@{}l*{7}{c}@{}} 
\toprule
\rotcell{} & 
\rotcell{Cam Ctrl} & 
\rotcell{Obj Ctrl} & 
\rotcell{3D Consist} & 
\rotcell{Photo Consist} & 
\rotcell{Motion Mag} \\ 
\midrule
$60\pm5$ over $30\pm5$          & {71.2\%} & {96.3\%} & {91.7\%} & {91.6\%} & {91.8\%} \\
$90\pm5$ over $60\pm5$       & {73.5\%} & {87.7\%} & {97.3\%} & {95.1\%} & {76.2\%} \\
\bottomrule
\end{tabular}%
\vspace{-0.2cm}
\caption{2AFC on \modelfull metrics with score difference 30.}
\vspace{-0.2cm}
\label{tab:re_30_60_90}
\end{table}

%% file: fig_text/re_resolution.tex
\begin{table}[t]
\resizebox{0.5\textwidth}{!}{%
\centering
\small 
\setlength{\tabcolsep}{3pt} 

\begin{tabular}{@{}l*{12}{c}@{}} 
\toprule
\rotcell{Res.} & 
\rotcell{Cam\\Ctrl} & 
\rotcell{Obj\\Ctrl} & 
\rotcell{Content\\Align} & 
\rotcell{3D\\Consist} & 
\rotcell{Photo\\Consist} & 
\rotcell{Style\\Consist} &
\rotcell{Subject\\Qual} &
\rotcell{Motion\\Acc} & 
\rotcell{Motion\\Mag} & 
\rotcell{Motion\\Smooth} \\ 
\midrule
1344×768          & {25.72} & {54.50} & {49.81} & {67.29} & {46.65} & {73.05} & {49.66} & {75.00} & {37.76} & {40.32} \\
256×256       & {25.69} & {53.78} & {50.32} & {67.41} & {47.06} & {73.88} & {48.99} & {74.89} & {36.90} & {39.62} \\
\bottomrule
\end{tabular}%
}
\vspace{-0.2cm}
\caption{\textbf{Robustness} to resolution and aspect ratio differences.}
\vspace{-0.2cm}
\label{tab:re_resolution}
\end{table}

%% file: supp_text/4_further_viz.tex
\section{Further Visualization}
\label{supp_sec:more_results}

Our \metric metrics provide a comprehensive assessment by decomposing the broad concept of ``world generation capability'' into 10 independent dimensions. The typical examples for each metric are presented in Figure~\ref{supp_fig:metrics_bigfigure} and Figure~\ref{supp_fig:metrics_bigfigure_dynamics}. Each row showcases the evaluation of a metric on two generated results, highlighting how \metric metrics effectively differentiate model performance.

We show performances of selected models on \metric-Dynamic in Figure~\ref{fig:radar_wsd} and \metric-Static in Figure~\ref{fig:radar_wss}. Figure~\ref{fig:radar_wsd} highlights the challenges that current video generation models face, with significant variations across different dimensions. Notably, all video generation models (\eg, Hailuo, VideoCrafter1-I2V, EasyAnimate, T2V-Turbo) exhibit very low \textit{camera controllability}, indicating difficulty in following predefined camera trajectories. Additionally, models (\eg, T2V-Turbo) that perform well in \textit{motion magnitude} tend to struggle with \textit{motion smoothness}, suggesting a trade-off between large movements and temporal stability.

In Figure~\ref{fig:radar_wss}, the evaluation of static world generation shows that 3D scene generation models (\eg, WonderWorld) achieve high \textit{camera controllability}, \textit{3D consistency} and \textit{photometric consistency}. However, they may struggle in \textit{subjective quality}, indicating that while they excel in maintaining geometric and photometric coherence, they may generate less visually appealing results.

\input{fig_text/supp_metrics_bigfigure}

\input{fig_text/supp_metrics_bigfigure_dynamics}

\input{fig_text/radar}

%% file: fig_text/supp_metrics_bigfigure.tex
\begin{figure*}[t]
    \centering
    \includegraphics[width=0.9\textwidth]{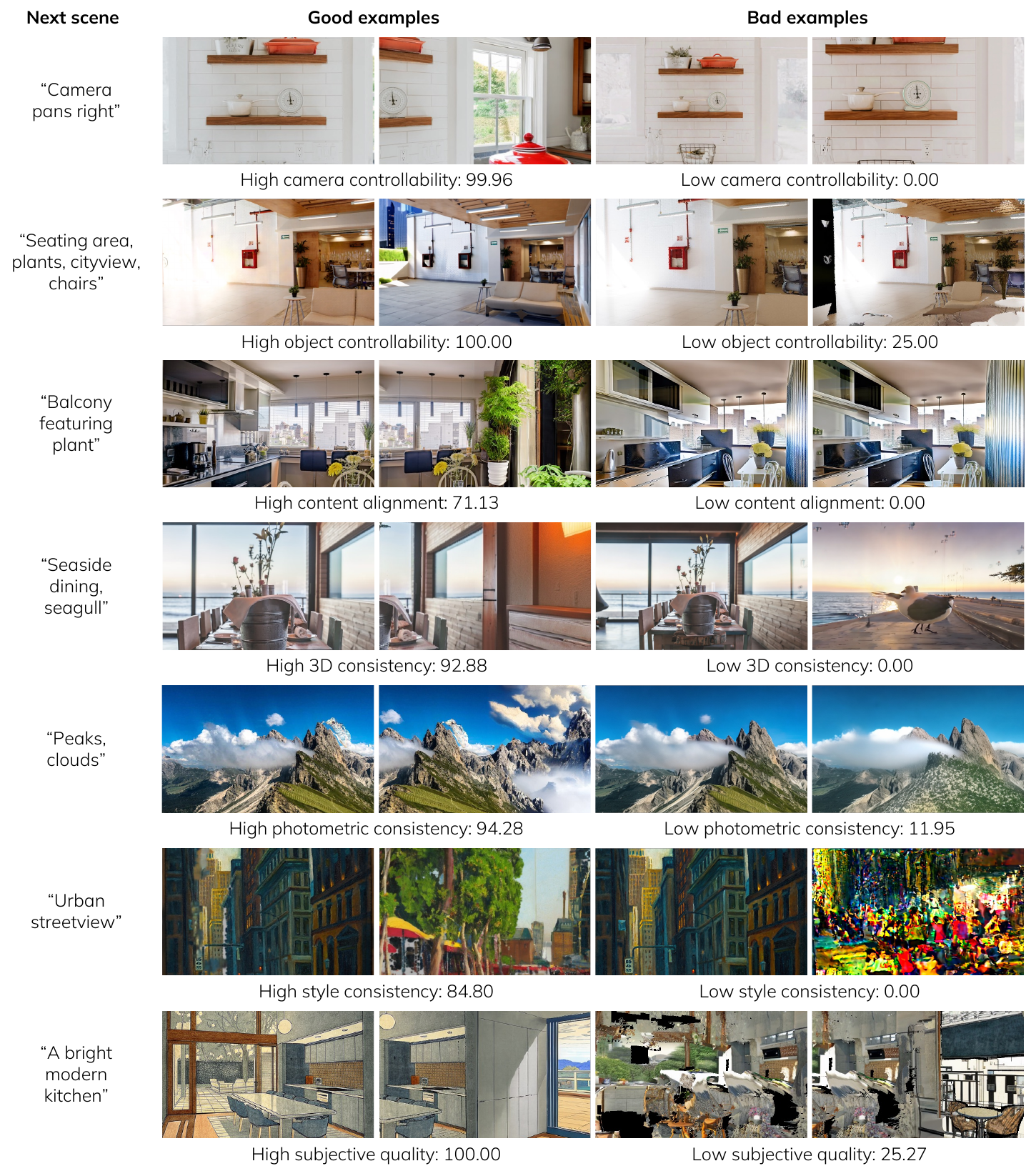}
    \caption{\textbf{Typical examples from controllability and quality aspects.} Each row showcases the evaluation of a metric on two generated results, where the good example is shown on the left, and the bad example is shown on the right.}
    \label{supp_fig:metrics_bigfigure}
\end{figure*}

%% file: fig_text/supp_metrics_bigfigure_dynamics.tex
\begin{figure*}[t]
    \centering
    \includegraphics[width=0.9\textwidth]{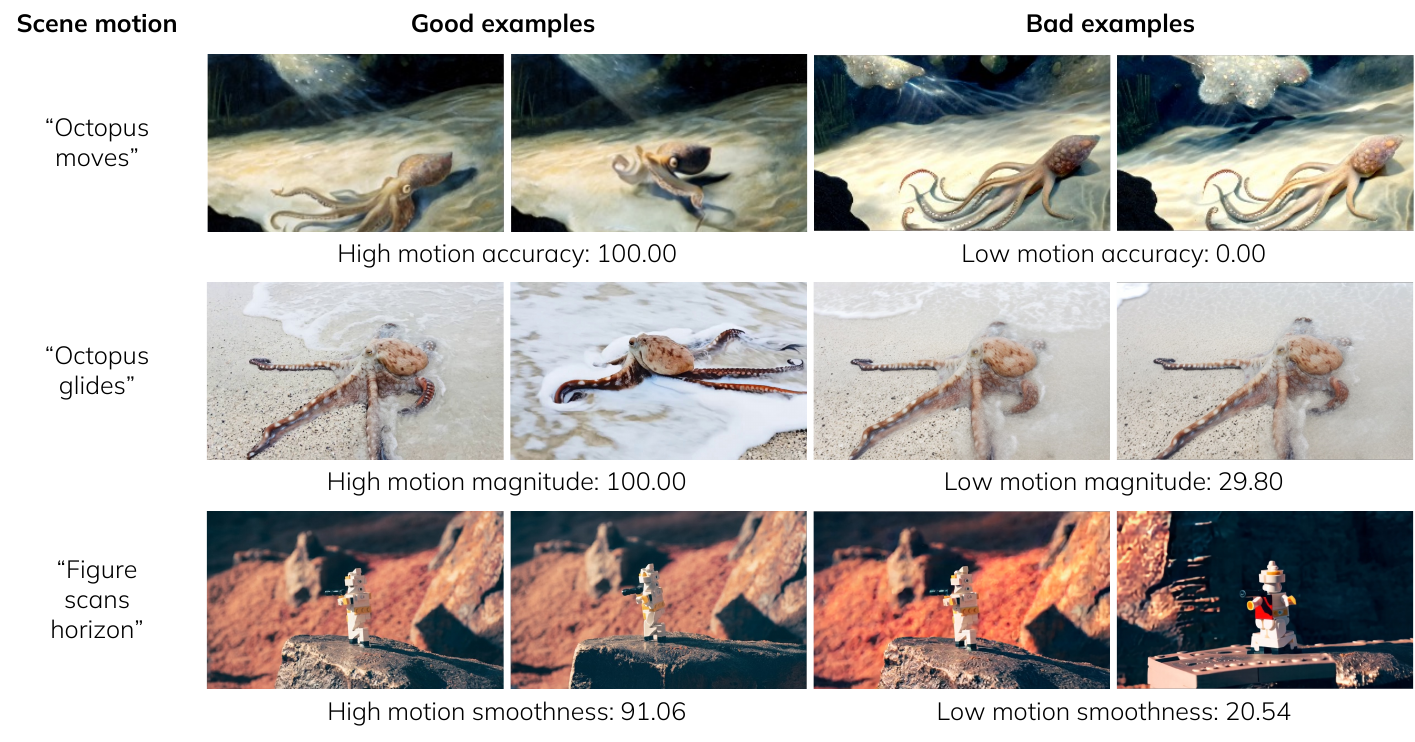}
    \caption{\textbf{Typical examples from dynamics aspect.} Each row showcases the evaluation of a metric on two generated results, where the good example is shown on the left, and the bad example is shown on the right.}
    \label{supp_fig:metrics_bigfigure_dynamics}
\end{figure*}

%% file: fig_text/radar.tex
\begin{figure*}[t]
    \centering
    \begin{minipage}{0.48\textwidth}
        \centering
        \includegraphics[width=\textwidth]{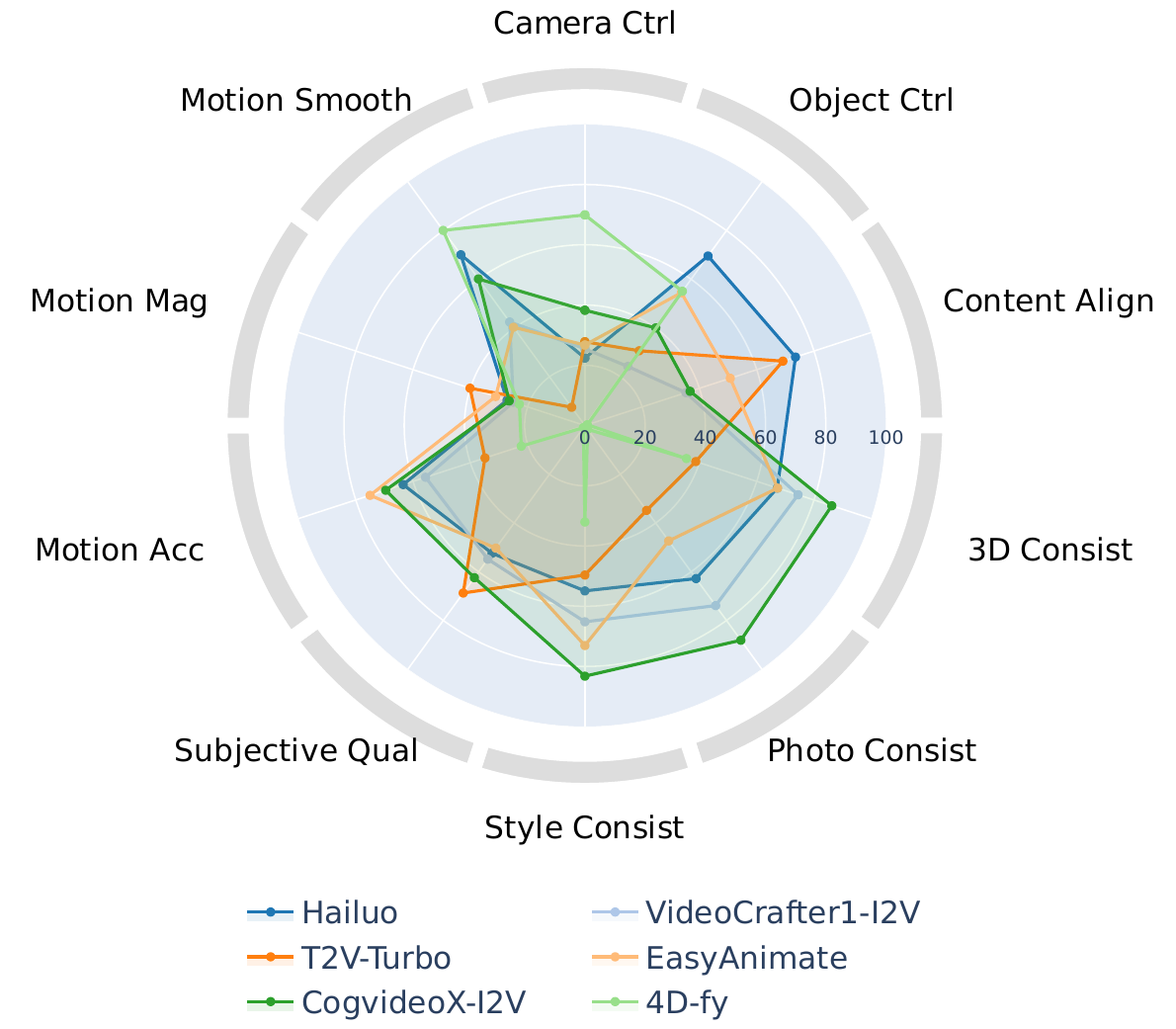}
        \caption{Evaluation results of \wsd on selected models.}
        \label{fig:radar_wsd}
    \end{minipage}
    \hfill
    \begin{minipage}{0.48\textwidth}
        \centering
        \includegraphics[width=\textwidth]{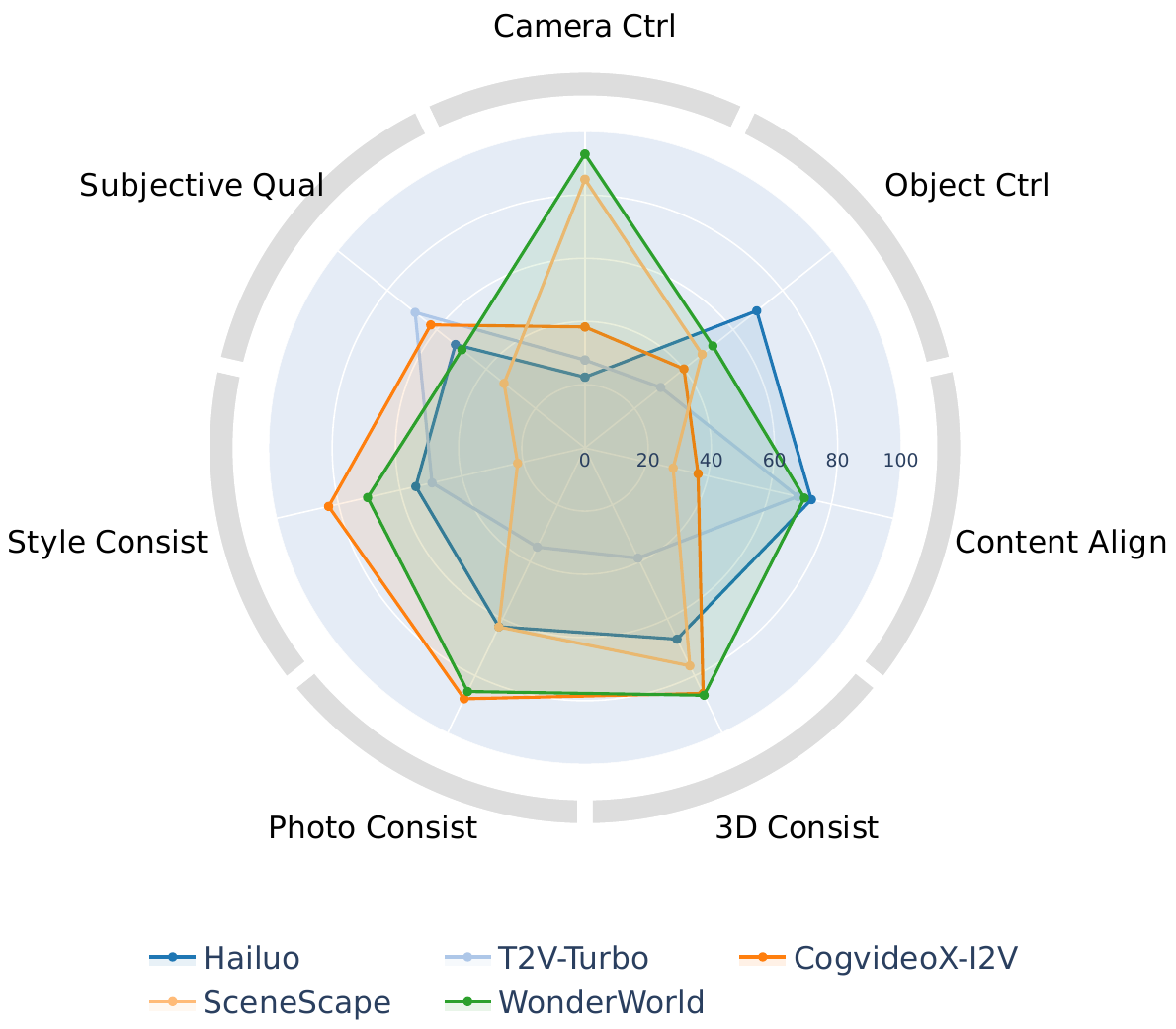}
        \caption{Evaluation results of \wss on selected models}
        \label{fig:radar_wss}
    \end{minipage}
\end{figure*}